\documentclass[a4paper,fleqn,usenatbib]{mnras}
\usepackage[T1]{fontenc}
\usepackage{ae,aecompl}

\usepackage{graphicx}	
\usepackage{amsmath}	
\usepackage{amssymb}	
\usepackage{graphics}	
\usepackage{booktabs}   
\usepackage{comment}
\usepackage{amssymb}
\usepackage{pifont}
\title[nIFTy V: Investigation of Infall Region]{nIFTy galaxy cluster simulations V: Investigation of the Cluster Infall Region}

\author[J. Arthur et al.]{
Jake Arthur$^{1}$\thanks{E-mail: jake.arthur@nottingham.ac.uk},
Frazer R. Pearce$^{1}$,
Meghan E. Gray$^{1}$,
Pascal J. Elahi$^{2,6}$,
\newauthor{
Alexander Knebe$^{3,4}$,
Alexander M. Beck$^{5}$,
Weiguang Cui$^{6,7}$,
Daniel Cunnama$^{8,9}$,
}
\newauthor{
Romeel Dav\'{e}$^{8,9,10}$,
Sean February$^{11}$,
Shuiyao Huang,$^{12}$,
Neal Katz,$^{12}$,
Scott T. Kay$^{13}$,
}
\newauthor{
Ian G. McCarthy$^{14}$,
Giuseppe Murante$^{15}$,
Valentin Perret$^{16}$,
Chris Power$^{6,7}$,
}
\newauthor{
Ewald Puchwein$^{17}$,
Alexandro Saro$^{18}$,
Federico Sembolini$^{3,4}$,
Romain Teyssier$^{19}$,
}
\newauthor{
Gustavo Yepes$^{3,4}$
}
\\
$^{1}$School of Physics \& Astronomy, University of Nottingham, Nottingham NG7 2RD, UK\\
$^{2}$Sydney Institute for Astronomy, A28, School of Physics, The University of Sydney, 
NSW 2006, Australia\\
$^{3}$Departamento de F\'{o}sica Te\'{o}rica, M\'{o}dulo 8, Facultad de Ciencias, 
Universidad Aut\'{o}noma de Madrid, 28049 Madrid, Spain\\
$^{4}$Astro-UAM, UAM, Unidad Asociada CSIC\\
$^{5}$University Observatory Munich, Scheinerstr. 1, D-81679 Munich, Germany\\
$^{6}$International Centre for Radio Astronomy Research, University of 
Western Australia, 35 Stirling Highway, Crawley,\\
 Western Australia 6009, Australia\\
$^{7}$ARC Centre of Excellence for All-Sky Astrophysics (CAASTRO)\\
$^{8}$Physics Department, University of the Western Cape, Cape Town 7535, South Africa\\
$^{9}$South African Astronomical Observatory, PO Box 9, Observatory, Cape Town 7935, South Africa\\
$^{10}$African Institute of Mathematical Sciences, Muizenberg, Cape Town 7945, South Africa\\
$^{11}$Center for High Performance Computing, CSIR Campus, 15 Lower Hope Street, 
Rosebank, Cape Town 7701, South Africa\\
$^{12}$Astronomy Department, University of Massachusetts, Amherst, MA 01003, USA\\
$^{13}$Jodrell Bank Centre for Astrophysics, School of Physics and Astronomy, 
The University of Manchester, Manchester M13 9PL, UK\\
$^{14}$Astrophysics Research Institute, Liverpool John Moores University, 
146 Brownlow Hill, Liverpool L3 5RF, UK\\
$^{15}$INAF, Osservatorio Astronomico di Trieste, via G.B. Tiepolo 11, I-34143 Trieste, Italy\\
$^{16}$Institute for Computational Science, ETH Zurich, Wolfgang-Pauli-Strasse 16, CH-8093, 
Zurich, Switzerland\\
$^{17}$Institute of Astronomy and Kavli Institute for Cosmology, University of Cambridge, 
Madingley Road, Cambridge CB3 0HA, UK\\
$^{18}$Department of Physics, Ludwig-Maximilians-Universitat, Scheinerstr. 1, 81679 Munchen, 
Germany\\
$^{19}$Institute of Theoretical Physics, Universit\"at Z\"urich, Winterthurerstrasse 190, 
CH-8057 Z\"urich, Switzerland\\
}

\date{Accepted XXX. Received YYY; in original form ZZZ}

\pubyear{2015}

\newcommand{\halos}{haloes}
\newcommand{\Halos}{Haloes}
\newcommand{\Fig}[1]{Fig.~\ref{#1}}

\def \gadget{{\sc GADGET}}

\def \arepo{{\sc A}repo}
\def \arepoIL{{\sc A}repo-{IL}}
\def \arepoSH{{\sc A}repo-{\sc SH}}

\def \gx{{\sc G3-X}}

\def \magneticum{{\sc G3-M}agneticum}
\def \pesph{{\sc G3-PESPH}}
\def \music{{\sc G3-MUSIC}}
\def \musicP{{\sc G3-MUSICP}i}
\def \owls{{\sc G3-OWLS}}

\def \gxx{{\sc G2-X}}

\def \ramses{{\sc RAMSES}}

\def \gimic{{\sc GIMIC}}

\begin{document}
\label{firstpage}
\pagerange{\pageref{firstpage}--\pageref{lastpage}}
\maketitle

\begin{abstract}
We examine the properties of the galaxies and dark matter \halos\ residing in the cluster infall region surrounding the simulated $\Lambda$CDM galaxy cluster studied by \citet{elahi2015} at $z=0$. The $1.1\times10^{15}h^{-1}\text{M}_{\odot}$ galaxy cluster has been simulated with eight different hydrodynamical codes containing a variety of hydrodynamic solvers and subgrid schemes. All models completed a dark-matter only, non-radiative and full-physics run from the same initial conditions. The simulations contain dark matter and gas with mass resolution $m_{\text{DM}}=9.01\times 10^8h^{-1}\text{M}_{\odot}$ and $m_{\text{gas}}=1.9\times 10^8h^{-1}\text{M}_{\odot}$ respectively. We find that the synthetic cluster is surrounded by clear filamentary structures that contain $\sim60\%$ of \halos\ in the infall region with mass $\sim 10^{12.5} - 10^{14} h^{-1}\text{M}_{\odot}$, including 2-3 group-sized \halos\ ($> 10^{13}h^{-1}\text{M}_{\odot}$). However, we find that only $\sim10\%$ of objects in the infall region are sub\halos\ residing in \halos, which may suggest that there is not much ongoing preprocessing occurring in the infall region at $z=0$. By examining the baryonic content contained within the \halos, we also show that the code-to-code scatter in stellar fraction across all halo masses is typically $\sim2$ orders of magnitude between the two most extreme cases, and this is predominantly due to the differences in subgrid schemes and calibration procedures that each model uses. Models that do not include AGN feedback typically produce too high stellar fractions compared to observations by at least $\sim1$ order of magnitude.
\end{abstract}




\section{Introduction}
In the $\Lambda$CDM paradigm galaxy clusters are built hierarchically by accreting smaller objects from the cluster infall region \citep{springel2005}, which we define here as the volume outside the galaxy cluster's virial radius. As galaxies fall into a cluster, their internal properties are significantly affected by their local environment, an effect that is more apparent nearer the overdense cluster centre \citep{dressler1980,lewis2002,gomez2003,hogg2004,poggianti2006,bamford2009}. Here, several physical mechanisms are thought to quench a galaxy's star formation or alter its morphology as it infalls (for review see \citet{boselli2006}). 

In the cluster centre it is difficult to disentangle these mechanisms, but by studying objects in the infall region we can not only examine what is building these clusters, but also possibly break this degeneracy. However, understanding cluster-specific phenomena is not the only reason to study the infall region of a galaxy cluster.  Many observational and theoretical studies have now raised the question of how important \textit{preprocessing} is, whereby some physical process is able to initiate significant changes as galaxies fall into groups and filaments well outside the virial region \citep{fujita2004,mcgee2009,bahe2013,cybulski2014}. However, preprocessing can be observationally difficult to study due to contamination from backsplash galaxies, which are galaxies that have already entered the cluster core, undergone significant disruption and travelled back out to the cluster outskirts. In fact, by using dark-matter-only simulations \citet{gill2005} found that $\sim50\%$ of galaxies residing between $R_{200}-2R_{200}$ of the main cluster halo are backsplash galaxies. 

Hydrodynamical simulations are now vital tools in aiding and interpreting astronomical observations of galaxy clusters \citep{borgani2011}, enabling us to track and quantify environmental effects as galaxies fall into the cluster. For example, \citet{bahe2015} used the \gimic\ simulations \citep{crain2009} to track galaxies falling into groups and clusters in order to understand the characteristic timescales of each environmental quenching mechanism and in what environment each dominated. Simulations are therefore invaluable for studying preprocessing in the cluster infall region, but before concrete conclusions can be drawn, the validity of simulations must be checked.

Hydrodynamical simulations model dark matter and gas coupled together through gravity, and evolve gas with the hydrodynamic equations. These equations are typically solved with either Langrangian Smoothed Particle Hydrodynamics (SPH) \citep{gingold1977,lucy1977,springel2010sph}, or Eulerian mesh-based schemes with optional Adaptive Mesh Refinement (AMR) \citep{stone1992,cen1992,kravtsov1997,teyssier2002}. The most famous comparison between state-of-the-art codes employing these numerical schemes was The Santa Barbara Cluster Comparison Project in \citet{frenk1999}. This study showed that mesh-based codes produced a simulated galaxy cluster with a cored entropy profile, which was worringly absent in the SPH codes. 

Since then more comparison studies have gone on to highlight other problems inherent in each numerical scheme. SPH methods typically have low shock resolution, poor accuracy in the treatment of contact discontinuities, and they have been shown to suppress fluid instabilities \citep{agertz2007}. In addition Eulerian mesh schemes are not strictly Galilean invariant, making the results sensitive to bulk velocities \citep{tasker2008}, which is particularly concerning for simulations of galaxy formation. More recently hybrid schemes and improved SPH schemes have been developed to account for these problems \citep{read2010,springel2010arepo,hopkins2014}.

On the other hand, the baryonic physics governing galaxy formation still remains uncertain, and including it complicates the simulations further. The focus has now shifted to creating simulations that are able to reproduce realistic galaxies \citep{vogelsberger2014,schaye2015}. The idea is to model the cooling and radiative physics that occurs as gas is converted into stars, and as feedback drives powerful outflows. More specifically, codes are now trying to model galaxy formation by including processes such as gas cooling \citep[e.g.][]{pearce2000,wiersma2009}, formation of stars from overdense gas \citep[e.g.][]{springel2003,schaye2008}, injection of energy from supernova \citep[e.g.][]{dalla2012}, growth of black holes \citep[e.g.][]{dimatteo2005}, and outflows from AGN accretion \citep[e.g.][]{booth2009}. Due to the large range of spatial and temporal scales that these mechanisms cover, they are simplified with analytical prescriptions containing tunable free parameters, namely \textit{subgrid} physics. These subgrid prescriptions still remain the largest uncertainties in galaxy formation simulations, with each simulation using their own preferred analytical prescriptions and calibrating the free parameters differently. 

The problems that plague modern galaxy formation simulations have prompted a rise in important comparison studies such as AQUILA and AGORA \citep{scannapieco2012,kim2014}. Projects such as these have investigated simulated galaxies resulting from different combinations of hydrodynamic solvers, subgrid schemes and resolution. This paper is a continuation of one such study, the \textit{nIFTy cluster comparison  project}. In the nIFTy cluster comparison series we use several different SPH and mesh codes, each equipped with their own preferred and calibrated subgrid schemes, to study the formation and evolution of a large $M_{200} = 1.1\times10^{15}h^{-1}\text{M}_{\odot}$ galaxy cluster produced by each code. The largest objects within the background dark matter distributions between all codes have been sufficiently aligned following a prescription described in paper I \citep{sembolini2016}, allowing a robust comparison to be carried out between hydrodynamic solvers and subgrid prescriptions included in each code. Also, by focusing on a simulated galaxy cluster, we can compare different codes in a variety of overdensities with a statistically robust sample of \halos. 

Due to recent improvements in SPH and mesh-based hydrodynamic solvers, the intial paper in the nIFTy series \citep{sembolini2016} revisted the work done in \citet{frenk1999} by examining the bulk properties of the simulated galaxy cluster at $z=0$ in both dark-matter-only and non-radiative (including gas but not cooling) runs. They found there was very good agreement in the dark matter density profiles between all codes, but the scatter in gas density profiles was of order a factor of $\sim2$. Most importantly, they found that the codes that employed a modern SPH scheme were able to reproduce an entropy core seen in the mesh based codes. 

Paper II \citep{sembolini2016b} analysed the effect the inclusion of full radiative baryonic physics had on the bulk properties of the simulated cluster at $z=0$. When including the uncertain baryonic physics, they found there is significantly more scatter in the bulk properties between codes in the full-physics run compared to the non-radiative run. The entropy profiles were also strongly affected by the radiative processes and washed out any differences between classic and modern SPH. Since then, \citet{cui2016} focused on the effect of including baryons on the galaxy cluster mass and kinematic profiles, as well as global measures of the cluster (e.g. mass, concentration, spin and shape). They found a good consistency ($\lesssim20$ per cent) between global properties of the cluster predicted by different codes when integrated quantities are measured within the virial radius $R_{200}$. However, there are larger differences for quantities within $R_{2500}$.

In paper III, \citet{elahi2015} (hereafter E16) analysed the sub\halos\ and galaxies produced by each code inside the central $1.8h^{-1}\text{Mpc}$ region surrounding the cluster. Whilst the code-to-code scatter in subhalo abundance was low in the dark-matter-only and non-radiative runs (codes differed by up to a factor of 1.3 and 1.9 respectively), the scatter was amplified in the full-physics run when the subgrid physics was included. Here codes differed by up to a factor of $\sim 2.4$. The discrepancy between codes in galaxy abundance is even worse: differences here extended up to a factor of 20 between the most extreme cases.

We would expect the code-to-code scatter in E16 to be mainly attributable to the different subgrid prescriptions and calibration methods each code uses. However, in the over dense centre differences in the gas environments are largest due to different hydro solvers and subgrid schemes between the models, and this could potentially have a sizeable effect on the code-to-code scatter seen in the central region. Therefore, this begs the question: in E16 do the differences in the subgrid schemes dominate the code-to-code scatter and how much is due to the different gas environments in which the \halos\ and galaxies live? To investigate this we have extended the work done in E16 by studying the simulated galaxy cluster infall region at $z=0$. By using objects within a sphere of radius $5h^{-1}\text{Mpc}$ centred on the cluster centre of mass, we have investigated whether the code-to-code scatter persists out to the less overdense infall region and how well each participating code can match to observed stellar and gas fractions. Also, by studying the infall region, we may investigate what is currently building our synthetic cluster.

The paper is organised as follows. In Section \ref{sec:methods} we briefly describe the participating codes, the simulated galaxy cluster and how we produced our halo catalogues. We present our results in Section \ref{sec:results}. Section \ref{sec:conclusions} contains a discussion along with our conclusions.

\section{Numerical Methods}\label{sec:methods}

\subsection{Codes}\label{sec:codes}
In this study we compare eight state-of-the-art hydrodynamical codes that contain calibrated subgrid physics. These include one Adaptive Mesh Refinement code, \ramses, the moving mesh code, \arepo, and 6 variants of the SPH code \gadget, \magneticum, \gx, \pesph, \music, \owls\ and \gxx. An extensive summary of how each code solves the hydrodynamic equations is presented in paper I of the nIFTy series. 

Each code incorporates their own preferred subgrid schemes for dealing with gas cooling/heating, star formation \&\ feedback, stellar population properties \&\ chemistry, and SMBH growth \&\ AGN feedback; the details of which are included in paper II and are also summarised in Table 1 in E16. For ease, we have also included a brief summary of the participating models in Table \ref{tab:codes}. We note that \ramses\ employs thermal AGN feedback and no stellar feedback to moderate cooling \citep{teyssier2002,teyssier2011}. \arepo\ has been run twice with variant subgrid physics, one including AGN feedback (\arepoIL) and one not including it (\arepoSH) \citep{vogelsberger2013,vogelsberger2014}. \arepoSH\ is not a production code, and has only been included in this study to observe the effect of switching off AGN feedback. \music\ includes no AGN feedback and only moderates cooling using stellar feedback based on \citet{springel2003} (hereafter SH03) \citep{sembolini2013}. A second variant of \music\ has been run, \musicP, with modified kinetic feedback described in \citet{piontek2011}. \pesph\ does not include AGN feedback, but uses a SH03 stellar feedback scheme with additional quenching in massive galaxies based on \citet{rafieferantsoa2015} (Huang et al. (in prep.)). \owls\ \citep{schaye2010}, \gxx\ \citep{pike2014}, \gx\ \citep{beck2016}, and \magneticum\ \citep{hirschmann2014} all employ some combination of stellar feedback and thermal AGN. 

\begin{table}
    \centering
     \begin{tabular}{cccc}
        \hline
	\hline
        Type & Model & SN & AGN \\
	\hline 
        \hline 
        AMR & \ramses & \ding{55} & \ding{51}     \\
        Moving Mesh & \arepoIL & \ding{51} & \ding{51}    \\
          & \arepoSH & \ding{51} & \ding{55}        \\
        Classic SPH & \music & \ding{51} & \ding{55}     \\
         & \musicP & \ding{51} & \ding{55}     \\
         & \owls & \ding{51} & \ding{51}    \\
         & \gxx & \ding{51} & \ding{51}    \\
        Modern SPH & \gx & \ding{51} & \ding{51}   \\
         & \pesph & \ding{51} & \ding{55}    \\
         & \magneticum & \ding{51} & \ding{51}    \\
        \hline
      \end{tabular}
    \caption{A brief summary of the models used in this study specifying which ones include stellar (SN) and AGN feedback. }
    \label{tab:codes}
\end{table}

\subsection{Data}\label{sec:data}
We use an $\text{M}_{200} = 1.1\times10^{15}h^{-1}\text{M}_{\odot}$ galaxy cluster drawn from the MUSIC-2 catalogue 
\citep{sembolini2013,sembolini2014,biffi2014}, which is a mass-limited sample of re-simulated 
\halos\ selected from the MultiDark dark-matter-only cosmological simulation \citep{prada2012}. The  MultiDark simulation contains $2048^3$ particles in a cube with side length $1h^{-1}\text{Gpc}$, where the chosen cosmology corresponds to the best-fitting $\Lambda$CDM model to WMAP7+BAO+SNI data with cosmological parameters taking the values $\Omega_m=0.27$, $\Omega_b=0.0.469$,  $\Omega_{\Lambda}=0.73$, $\sigma_8=0.82$, $n=0.95$ and $h=0.7$ \citep{komatsu2011}. All the data from the MultiDark simulation is publicly available online through the MultiDark database \footnote{\url{https://www.cosmosim.org/}}.

The MUSIC-2 cluster catalogue was constructed by selecting all objects in the MultiDark volume with mass $>10^{14}h^{-1}\text{M}_{\odot}$ at $z=0$. These \halos\ were then resimulated using a zooming technique described in \citet{klypin2001}. In a low resolution ($256^3$) MultiDark volume, particles in a sphere of radius $6h^{-1}\text{Mpc}$ around each selected object were mapped back to their initial conditions. These initial conditions from the original simulations were then generated on a $4096^3$ size mesh, improving the mass resolution of the resimulated \halos\ by a factor of 8. Each code completed a Dark-Matter-only (DM), Non-Radiative (NR) and Full-Physics run (FP). The mass resolution of particles in the particle-based codes in the dark-matter-only simulations is $m_{\text{DM}}=1.09\times 10^9h^{-1}\text{M}_{\odot}$ and in the gas runs is $m_{\text{DM}}=9.01\times 10^8h^{-1}\text{M}_{\odot}$ and $m_{\text{gas}}=1.9\times 10^8h^{-1}\text{M}_{\odot}$. The grid resolution in the mesh codes was chosen to match these particle resolutions as shown in \citet{sembolini2016}.

\begin{figure*}
	\includegraphics[width=\textwidth]{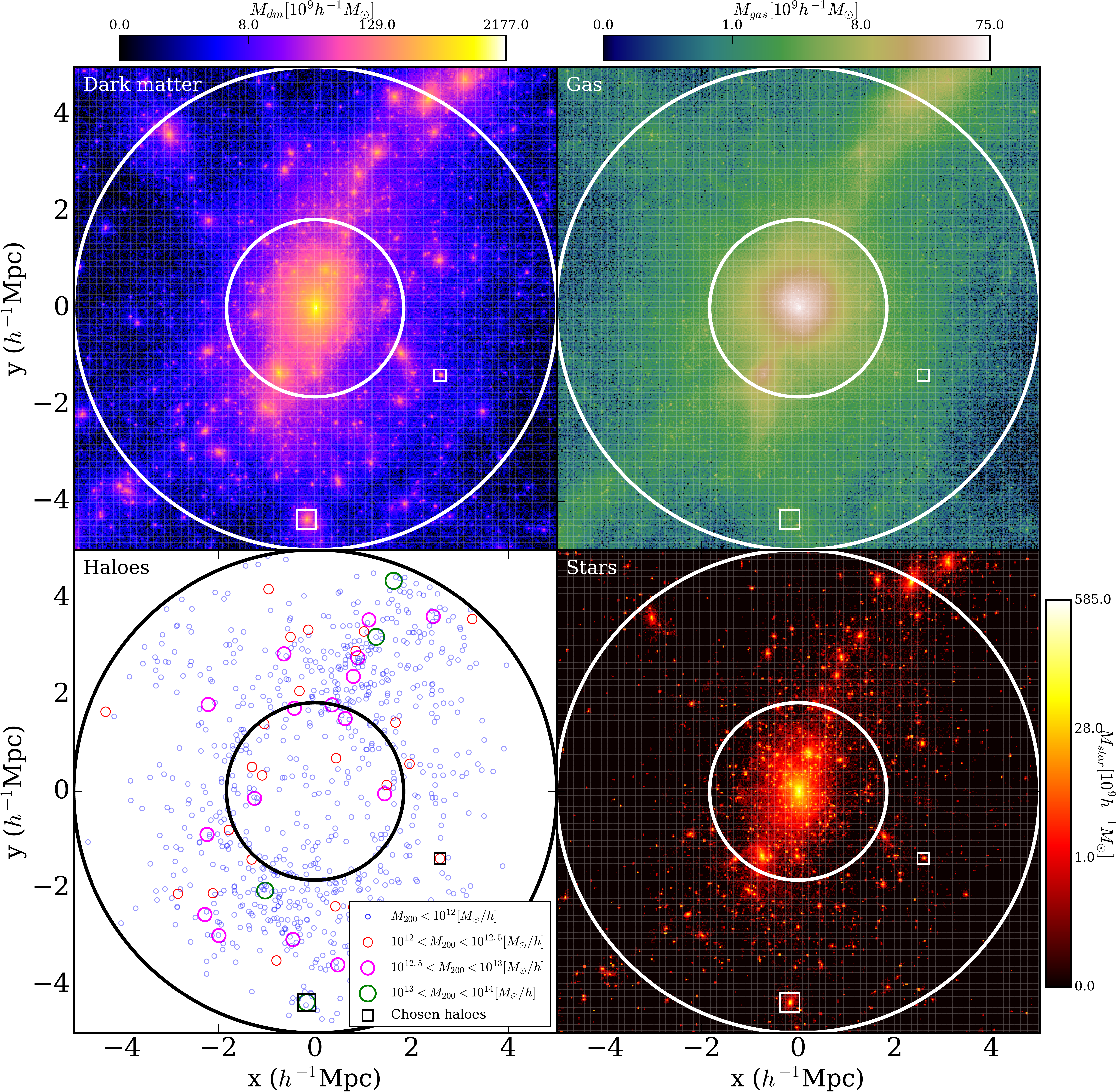}
    \caption{The distribution of dark matter (top left panel), gas
      (top right panel), stars (bottom right panel) and \halos\ (bottom
      left panel) for the \owls\ full-physics simulation. The colorbars are on a log scale. Each panel
      is $10 h^{-1}{\rm Mpc}$ across. A circle of radius $5 h^{-1}{\rm
        Mpc}$ is shown as a bold line on each panel. The inner circle
      indicates $R_{200}$ for the \music\ simulation, which is used to
      delineate the central cluster region from the infall region that
      lies between the two circles. The circles marked on the bottom
      left panel indicate the location of \halos\ or sub\halos\ and
      are colour-coded by mass as indicated in the legend. The black
      squares highlight the isolated \halos\ used for analysis in
      \Fig{fig:onetoone}. These chosen \halos\ are also indicated on
      the other three panels with small white squares.}
    \label{fig:clusterregion}
\end{figure*}

\begin{figure*}
	\includegraphics[width=\textwidth]{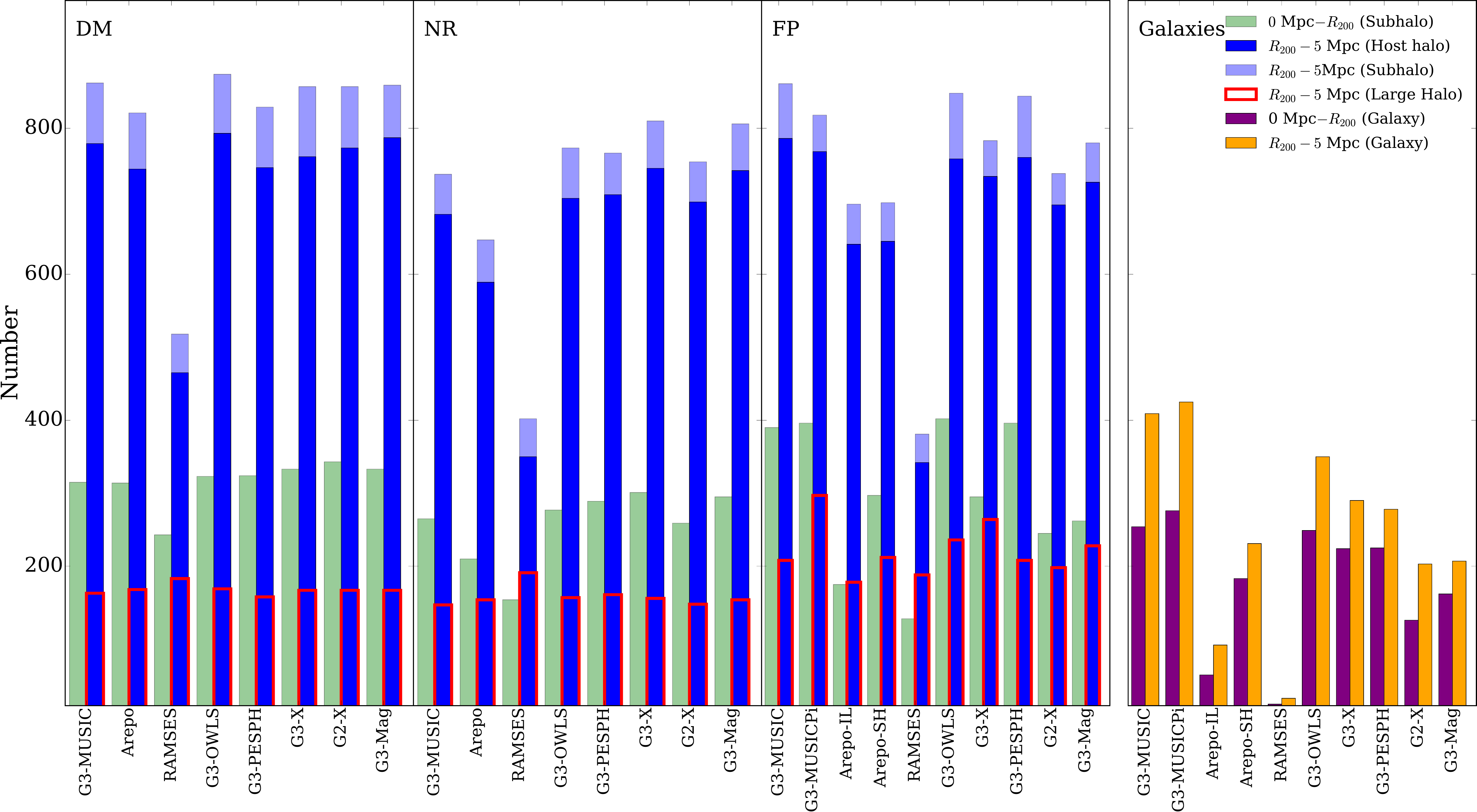}
    \caption{Number of \halos, sub\halos\ and galaxies. The left,
      left-centre and right-centre panels show the number of \halos\ and
      sub\halos\ for the dark-matter only, non-radiative and full-physics runs respectively.  The right panel shows the number of
      galaxies produced by each code in the full-physics run. In the first three
      panels, the green and blue bars represent different regions. The
      green bars are for the central virialised region inside
      $R_{200}$ of the reference \music\ simulation. The blue bars are
      the infall region, between $R_{200}$ and $5 h^{-1}\,{\rm Mpc}$.  Solid
      bars represent \halos, whilst transparent bars stacked on top
      represent sub\halos. The red outline indicates the number of
      large \halos, containing $\gtrsim 200$ or more particles. In the last panel the purple and orange bars represent galaxies in the central and infall regions respectively. For all simulations
      there are $\sim 10$ times more \halos\ than sub\halos\ in the infall
      region (blue), whilst in the central region (green) all but one
      of the objects are sub\halos.}
    \label{fig:numhalosandgals}
\end{figure*}

\subsection{Analysis}\label{sec:analysis}
\subsubsection{Halo Catalogues}
All \halos\ and sub\halos\ were identified and analysed using {\sc VELOCIraptor} \cite[aka {\sc stf}][freely available \href{https://github.com/pelahi/VELOCIraptor-STF.git}{\url{https://github.com/pelahi/VELOCIraptor-STF.git}}]{elahi2011}, which identifies \halos\ using a 3D Friends-Of-Friends (FOF) algorithm and then identifies sub\halos\ using a phase-space FOF algorithm. In this paper, a subhalo is a self-bound satellite object within the virial radius of another larger halo. Both \halos\ and sub\halos\ are indentified by only considering dark matter particles. {\sc VELOCIraptor} identifies selfbound structures as \halos\ or sub\halos\ once they contain a minimum of 20 particles. In our simulations, bound baryonic particles are associated with the halo or subhalo of the closest dark matter particle. As in \citet{elahi2015}, a galaxy in this study is defined as any self-bound structure that contains 20 or more star particles, corresponding to a galaxy mass of $\sim 2\times 10^{9}h^{-1}\text{M}_{\odot}$. 

\subsubsection{Contaminants Removal}
In this paper we study all objects within a sphere of radius
$5h^{-1}\text{Mpc}$ centred on the cluster centre of mass at $z=0$. As this is a zoom
simulation with a nested heirarchy of progressively lower mass
resolution shells, it is possible for low resolution dark matter
`interloper' particles to enter into the region of interest from the
low resolution outskirts.  We have traced these particles, and in all
of our simulations we find $\sim 20$ interloper particles in the
infall region, lying in two groups. We have removed all of the
\halos\ lying within $1h^{-1}\text{Mpc}$ of these groups from our
analysis. Only $\sim 30$ \halos\ are excluded using this approach, so even if we included them in any further analysis we do not expect them to cause any significant statistical changes.

\section{Results}\label{sec:results}

\subsection{\Halos\ and  Galaxies}\label{sec:halos}

We begin our analysis by first presenting the cluster produced by \owls\ in \Fig{fig:clusterregion}. The top-left, top-right and bottom-right panels show the projected density of dark matter, gas and stars across a $10h^{-1}\text{Mpc}$ square centred on the cluster respectively. Henceforth we define the `central' region of the cluster as the spherical volume contained within the inner circle, which is $R_{200}$ ($1.8h^{-1}\text{Mpc}$) of the central halo in the \music\ reference simulation. The difference in $R_{200}$ between the DM, NR and FP runs is $\lesssim 2\%$ \citep{cui2016}. We also define the `infall' region as the shell between the inner and outer circles, where the latter defines the (somewhat arbitrary) $5h^{-1}\text{Mpc}$ ($\sim 3 R_{200}$) boundary in this paper. The last panel shows the \halos\ existing only in the infall region (the \halos\ in the central region are not plotted here). Any \halos\ residing within the central region in \Fig{fig:clusterregion} are foreground objects.

There is clear filamentary structure surrounding the cluster at $z=0$, with two particularly dense filaments running towards the bottom-left and top-right regions of each panel. In order to see how the most massive group-sized \halos\ are distributed in the infall region, we have partitioned the \halos\ into four mass bins, shown as different sizes and colours. After a 3D inspection we find that $\sim 60\%$ of \halos\ with mass $\sim 10^{12.5} - 10^{14} h^{-1}\text{M}_{\odot}$ reside within filamentary structure at $z=0$, including 2-3 group sized ($\gtrsim 10^{13} h^{-1}\text{M}_{\odot}$) \halos.

Our first code-to-code comparison in this study is presented in \Fig{fig:numhalosandgals} where we show the number of \halos, sub\halos\ and galaxies produced by each participating code.  \arepoSH\ and \musicP\ only differ from their original variants in the FP run due to their variant subgrid prescriptions, so no values are shown for these codes in the DM and NR runs.

\Fig{fig:numhalosandgals} shows that nearly all codes produce a consistent number of \halos\ and sub\halos\ in both the infall and central regions in all runs, though there is more code-to-code scatter in the NR and FP runs due to the inclusion of uncertain baryonic physics. The exception is \ramses, which produces nearly a factor of two fewer objects than the median in the infall region across all runs. However, when we consider the large \halos\ in the infall region that have a minimum of 200 dark matter particles (red edged bars), we see that the codes are more consistent with each other across all runs, even \ramses. This suggests that \ramses\ is not resolving \halos\ that contain less than $\sim200$ particles, which has been shown before in AMR codes \citep{oshea2005}. In this instance, \ramses\ probably just needs to use a mesh with better resolution in order to resolve the smaller objects.

All codes produce $\sim 10$ times more \halos\ (solid blue bars) than sub\halos\ (transparent blue bars stacked on top) in the infall region across all runs, whilst nearly all objects in the central region are sub\halos\ residing within $R_{200}$ of the main cluster halo. The lack of sub\halos\ in the infall region indicates that in this cluster at $z=0$ our halo sample is not heavily contaminated by sub\halos\ currently undergoing some preprocessing. Dark-matter-only simulations produce similar subhalo to halo ratios, for example \citet{klypin2011} showed that in the Bolshoi simulation the ratio between subhalo and halo abundances is typically $\sim 10-20\%$ for halo masses between $\sim10^{9}h^{-1}\text{M}_{\odot} - 10^{14}h^{-1}\text{M}_{\odot}$. The low number of sub\halos\ that surround the cluster at $z=0$ may at first appear in tension with recent observational studies that have suggested preprocessing is a dominant mechanism at $z\sim0$ \citep{cybulski2014,just2015}. However, we should note that this may not be a fair comparison and we intend to carry out a full temporal study to investigate preprocessing as this cluster forms in future work.

E16 showed that there was a large inconsistency between codes in how many galaxies they produced within the central $2h^{-1}\text{Mpc}$ region, the scatter between codes extended up to a factor of $\sim 20$. Whilst \Fig{fig:numhalosandgals} corroborates this, the most notable result is that this code-to-code scatter persists out to the infall region as well, suggesting that it may not be the different gas environments driving the code-to-code scatter, but the different subgrid schemes each code employs. In the infall region \music\ and \musicP\ produce the most galaxies, which is expected as these two codes do not include AGN feedback and only moderate gas cooling with stellar feedback. \arepoIL\ and \ramses\ produce a factor of $\sim 3$ and $\sim 13$ fewer galaxies than the median respectively, a potential consequence of powerful AGN feedback tuned to match the properties of the central halo, which is quenching smaller objects very efficiently. We are confident that the scatter in galaxy abundances between codes here is not due to poorly resolved galaxies, as we see the code-to-code scatter extends up to a factor of $\gtrsim 25$ for well resolved galaxies as well ($\text{M}_{200} \gtrsim 10^{10}h^{-1}\text{M}_{\odot}$) as seen later in the text.

\begin{figure*}
    \centering
    \includegraphics[width=0.3605\textwidth]{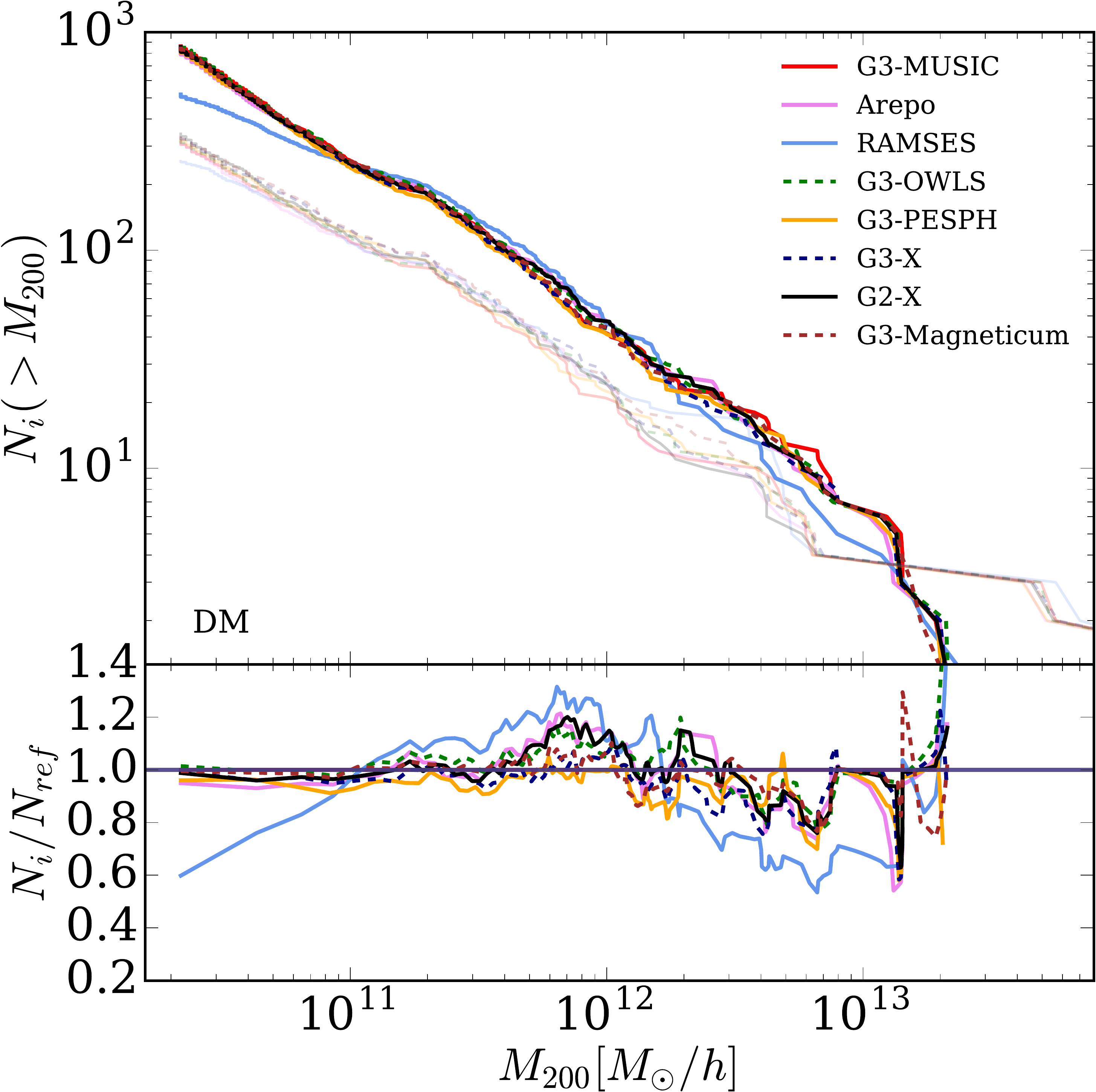}
    \includegraphics[width=0.313\textwidth]{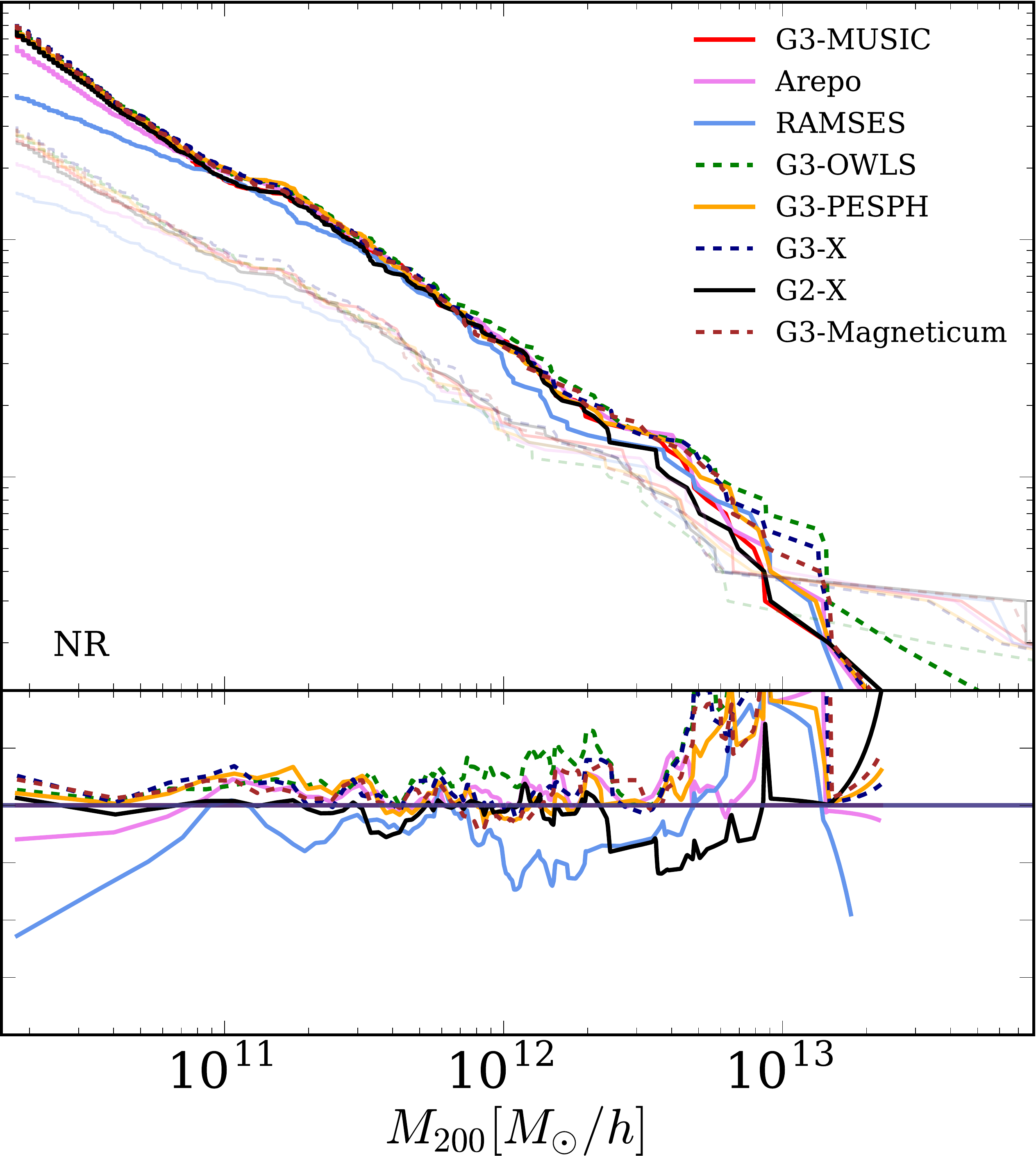}
    \includegraphics[width=0.313\textwidth]{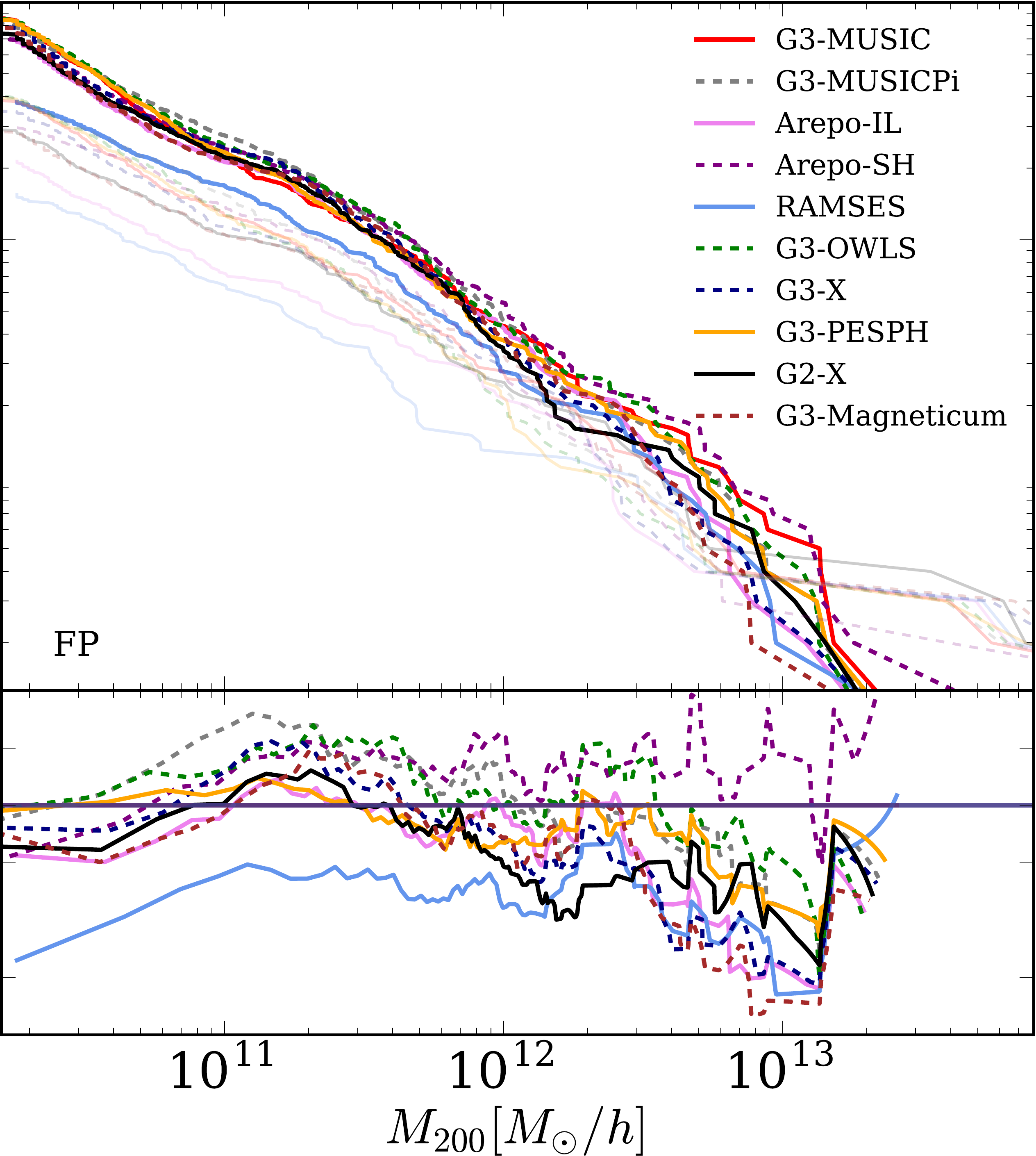}
    \caption{The top panels show the cumulative halo (including subhalo)
      mass functions for each simulation only considering the dark matter component of the objects, while the bottom panels show
      the ratio between each simulation and the reference model,
      \music. The left, centre and right panels show the results from
      the dark-matter only, non-radiative and full-physics runs
      respectively. Transparent and opaque lines represent the central
      and infall regions respectively. The transparent lines have only been included to show how the code-to-code scatter in the central region compares to the infall region; see \citet{elahi2015} for detail about the central region. The infall region contains more than
      twice as many \halos\ as the central region. \ramses\ is an outlier even
      for the dark-matter-only run in the infall region, with many
      small \halos\ missing. These missing \halos\ extend to around
      $10^{12} h^{-1}M_{\odot}$ (over 1000 particles) in the full
      physics run where the total number of \halos\ present in the
      infall region is around $40\%$ of that seen in the other
      models.}
    \label{fig:cummass}
\end{figure*}

\begin{figure*}
    \centering
    \includegraphics[width=0.3605\textwidth]{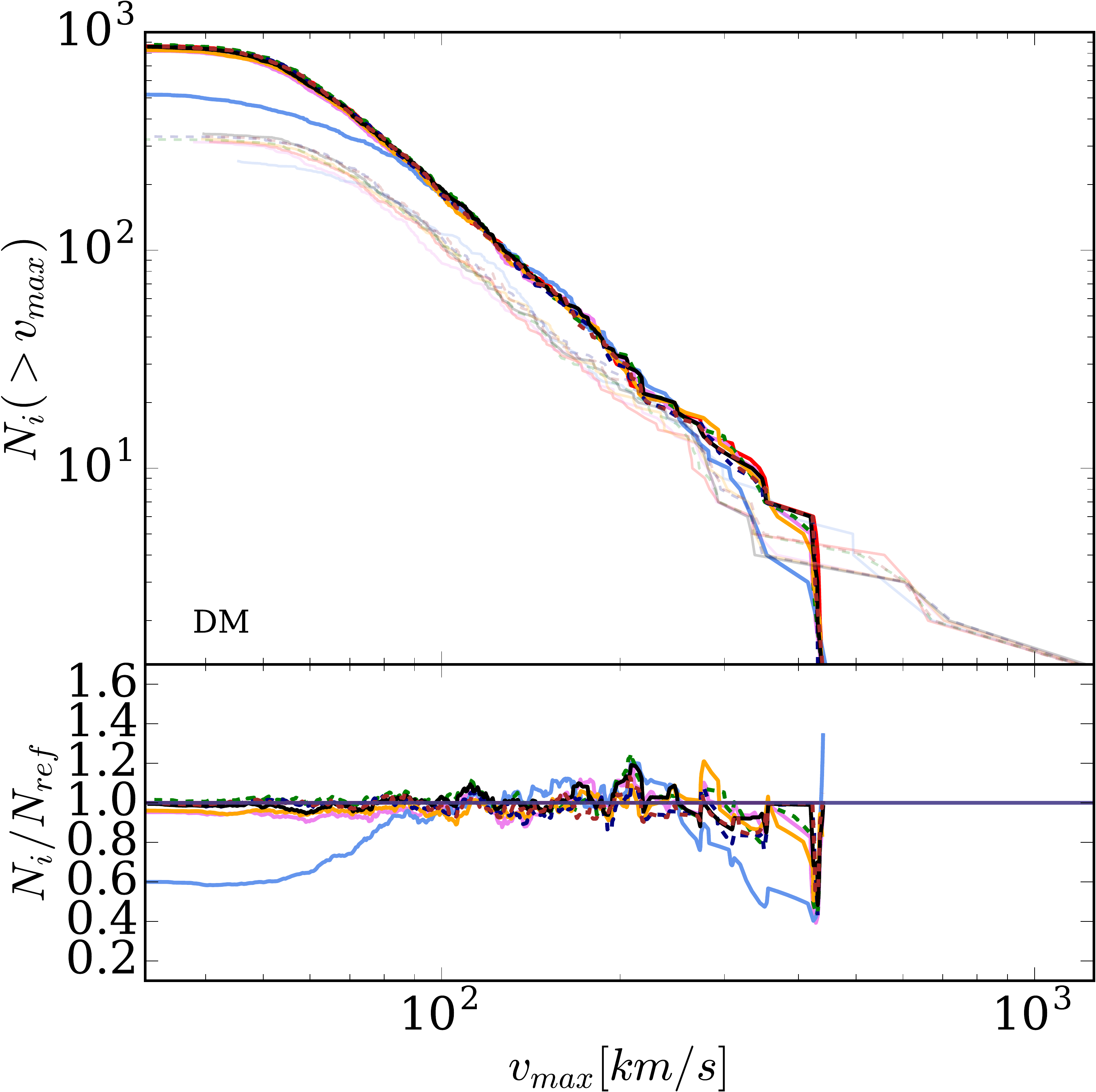}
    \includegraphics[width=0.313\textwidth]{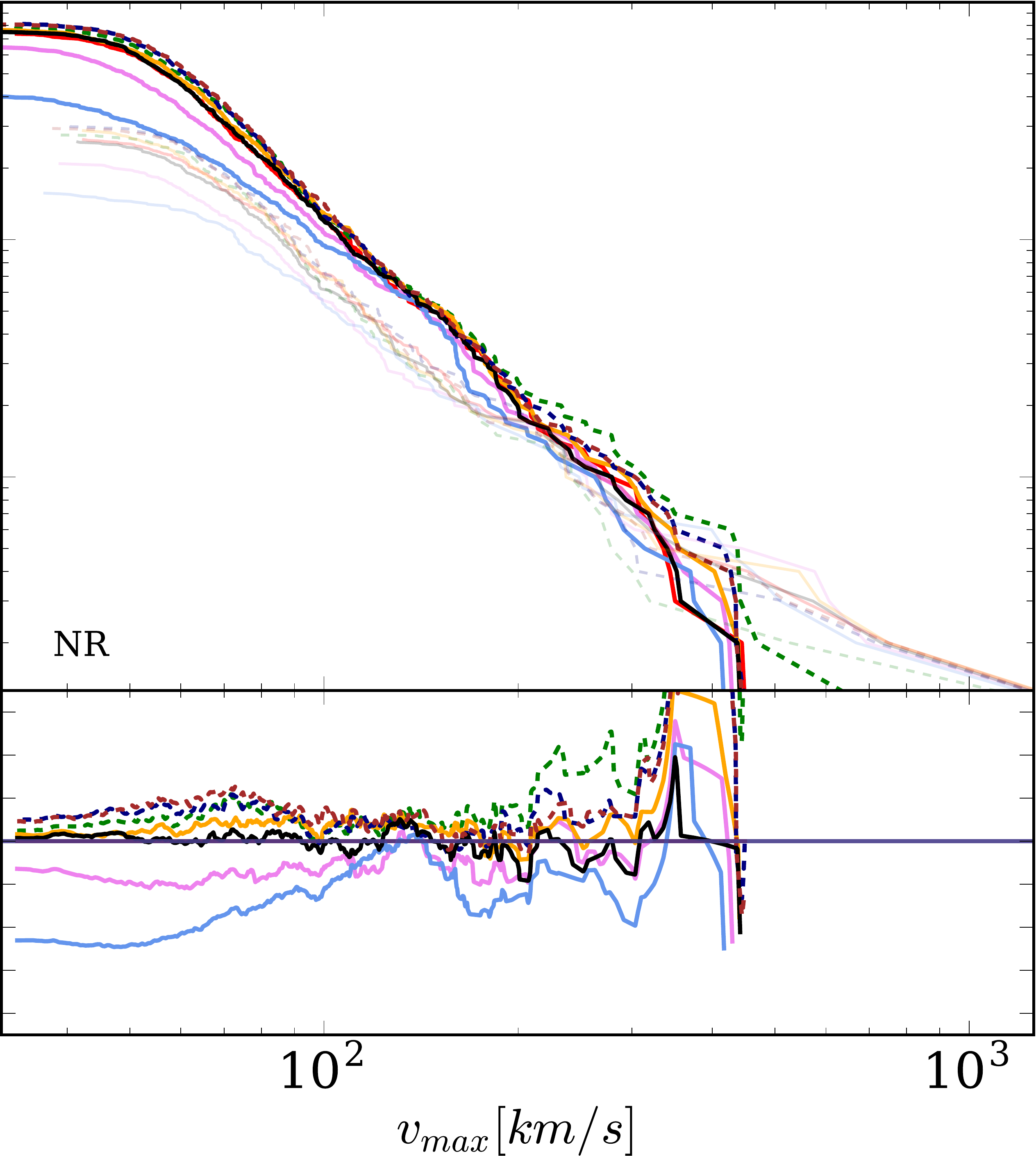}
    \includegraphics[width=0.313\textwidth]{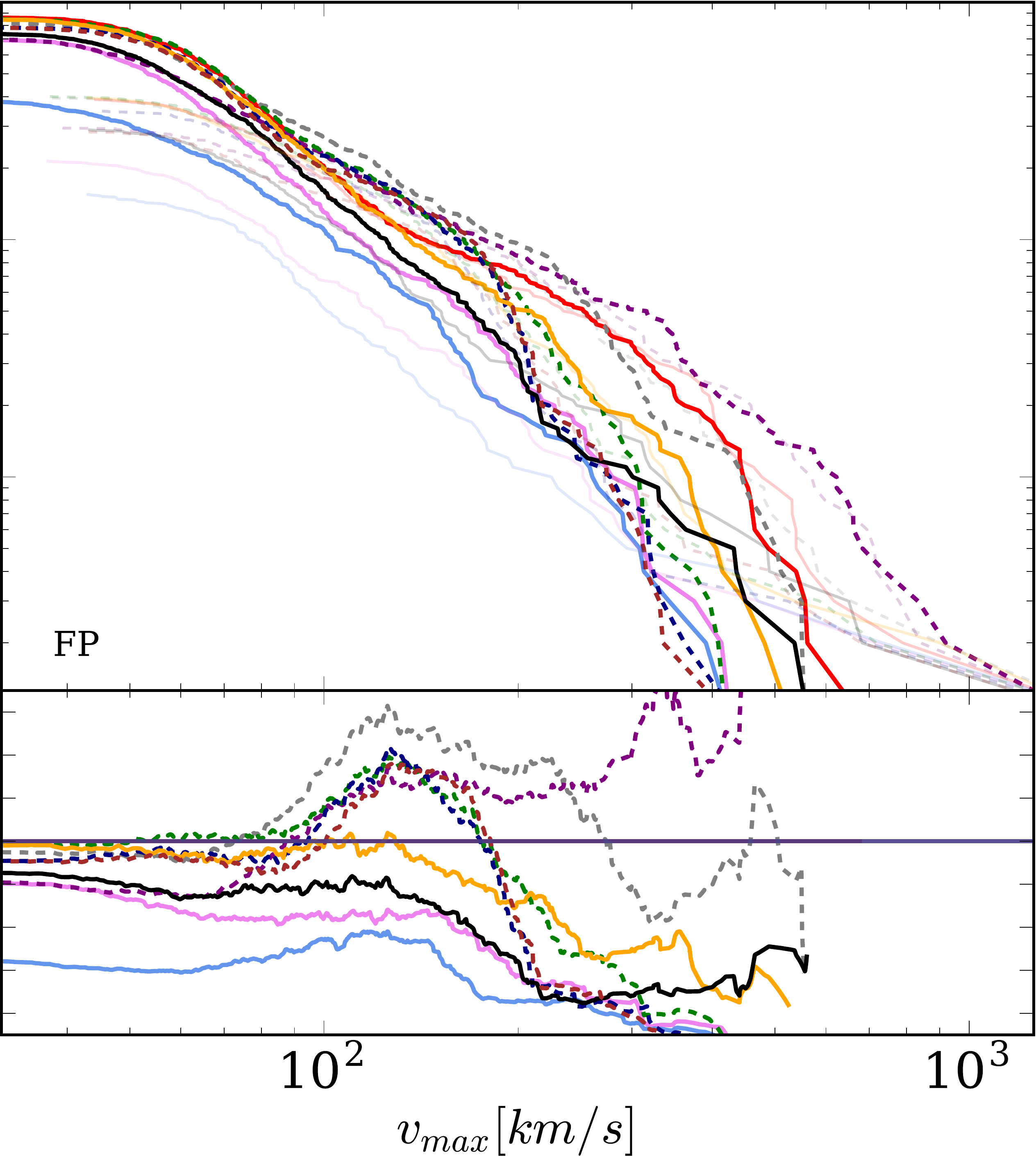}
    \caption{Similar to \Fig{fig:cummass}, but for the
      cumulative maximum circular velocity distribution  (see Figure \ref{fig:cummass} for
      legend). The plots
      show similar results to \Fig{fig:cummass}, though here in the FP run the code-to-code scatter is amplified compared to the corresponding cumulative mass function.}
    \label{fig:cumvmax}
\end{figure*}

We next investigate the mass functions and circular velocity distributions of \halos\ and sub\halos, shown in \Fig{fig:cummass} and \Fig{fig:cumvmax} respectively. A value for $\text{M}_{200}$ can be calculated for sub\halos\ in a similar fashion to \halos, however when $\text{R}_{200}$ cannot be found, $\text{M}_{200}$ is set to equal the total mass of the subhalo. We have displayed these distributions for both the central (transparent) and infall regions (opaque). The mass functions for the central region have only been included to show how the code-to-code scatter in the central region compares to the infall region; see \citet{elahi2015} for more detail about the central region. The lower panels in these figures show the residuals of these distributions in the infall region relative to the \music\ reference simulation. In the infall region the codes produce a largely consistent set of mass functions (\Fig{fig:cummass}, solid lines) in the DM run, where the typical scatter is $\lesssim 10\%$. As found in E16, we note that this scatter is increased in the NR run to $\lesssim 15\%$, because of the inclusion of gas and the different hydrodynamic approaches each code uses to evolve the gas particles. The code-to-code scatter is then amplified in the FP run to typically $\sim 60\%$ for all \halos, with the addition of uncertain subgrid effects. All codes produce twice as many \halos\ and sub\halos\ with mass $\lesssim 10^{12}h^{-1}\text{M}_{\odot}$ in the infall region compared to the centre across all runs. In total there are $\sim 3$ times as many \halos\ and sub\halos\ in the infall region ($\sim 900$ objects) compared to the centre ($\sim 300$ objects), which allows us to utilise a statistically robust sample of objects for this study.

In the DM run \ramses\ produces $\sim 40\%$ fewer \halos\ and sub\halos\ with mass $\lesssim 10^{11}h^{-1}\text{M}_{\odot}$ compared to all other codes in the infall region, a number which is consistent with \Fig{fig:numhalosandgals}. This is amplified in the FP run, where \ramses\ produces $\sim 50\%$ fewer \halos\ with mass $\lesssim 10^{12}h^{-1}\text{M}_{\odot}$ compared to most other codes. It is clear that the combination of absent low mass \halos\ and powerful AGN feedback has a dramatic effect on even quite large \halos\ for \ramses, impacting their number even for \halos\ that contain several thousand particles.

As the recovered mass (in this case $\text{M}_{200}$) is not observable, in \Fig{fig:cumvmax} we present the maximum circular velocity distributions. As \citet{knebe2011} demonstrated, these are less susceptible to outer boundary issues but require more particles to measure reliably and are known to be sensitive to central concentrations of sub\halos\ \citep{onions2013}. In \Fig{fig:cumvmax} nearly all codes are in good agreement in the DM and NR runs, but the underproduction of low mass \halos\ by \ramses\ and to some extent \arepo\ in the NR run is even more apparent. The most notable change in the maximum circular velocity distributions is the significant increase in code-to-code scatter in the FP run compared to the corresponding mass function. Typically the scatter in the FP mass function is $\sim 60\%$, whilst in the FP circular velocity distribution the code-to-code scatter extends up to $\sim 100-150\%$.  Clearly, the additional physics contained in the FP runs influences the central regions which are being probed by the measurement of the maximum circular velocity and this could be problematic for this approach. Interestingly, we find that the code-to-code scatter in the FP circular velocity distribution reaches a factor of more than two at $v_{max} \sim 200 kms^{-1}$, which corresponds to a halo mass $\sim 5\times10^{12}h^{-1}\text{M}_{\odot}$. It is clear that this scatter is not due to poorly resolved \halos, but more likely the internal subgrid prescriptions.

\begin{figure}
    \centering
    \includegraphics[width=\columnwidth]{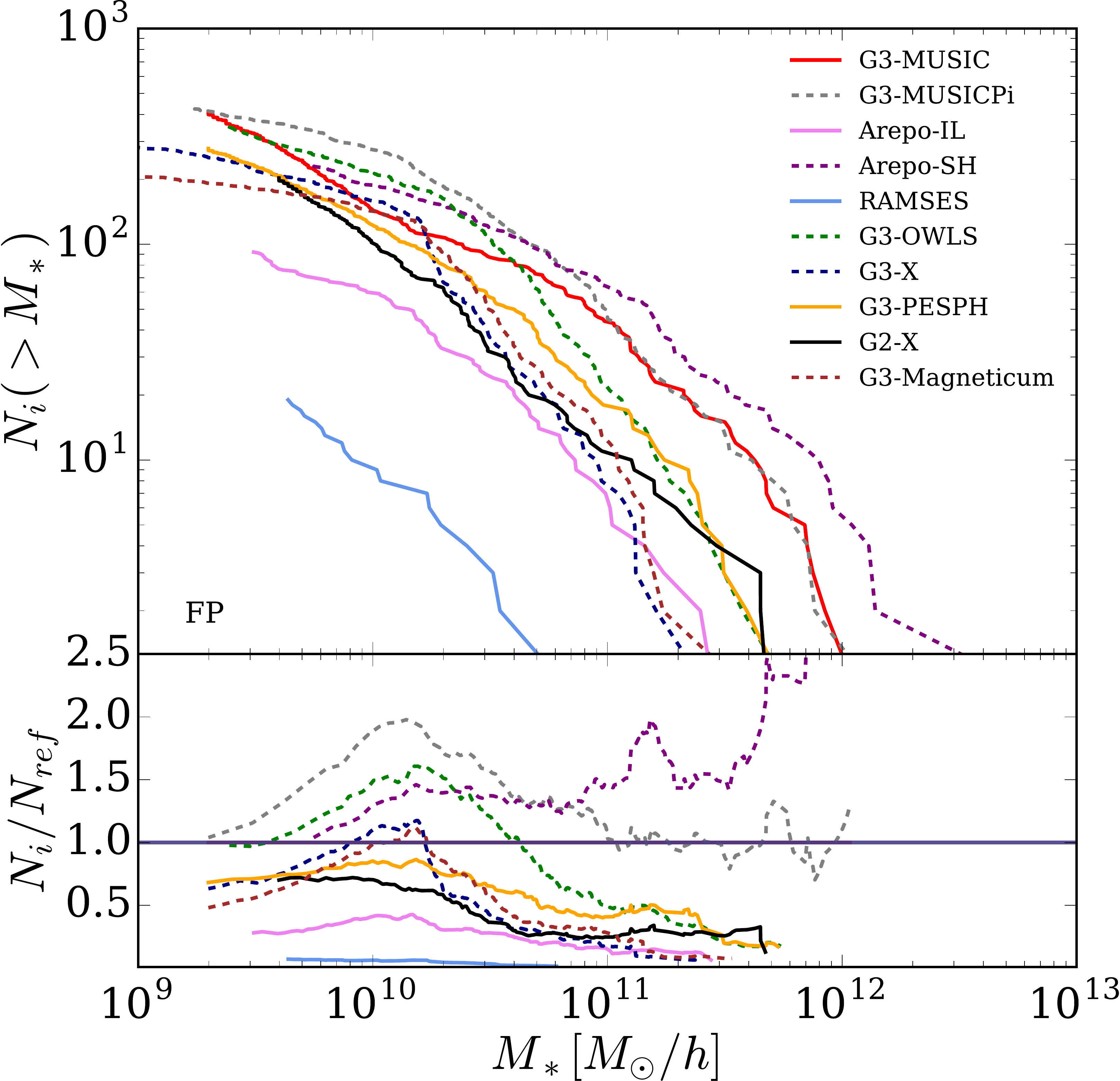}
    \caption{The top panel shows the cumulative galaxy stellar mass
      distribution in the cluster infall region, whilst the bottom
      panel shows the ratio of each code relative to the
      \music\ simulation. Even above $10^{11}h^{-1}\text{M}_{\odot}$ the code-to-code scatter extends beyond $\sim100\%$.}
    \label{fig:galcummass}
\end{figure}

In \Fig{fig:galcummass} we present the cumulative Galaxy Stellar Mass Function (GSMF) of galaxies in the cluster infall region produced by each code. The top panel shows the cumulative distribution, whilst the bottom panel shows the ratio of each GSMF with the GSMF produced by the \music\ reference simulation. The most notable result shows that above $10^{10}h^{-1}\text{M}_{\odot}$, where the galaxies are well resolved (these galaxies will contain $\gtrsim 100$ star particles), the scatter between the codes is of order $\sim100\%$.

The inability of \ramses\ to resolve small \halos\ coupled with the fact that it employs a powerful AGN feedback scheme causes the code to produce no galaxies above $\sim 10^{10.6}h^{-1}\text{M}_{\odot}$, and below this mass \ramses\ produces an order of magnitude fewer galaxies compared to the other codes. Conversly, \arepoSH\ produces the most massive galaxies primarily because it does not include an AGN feedback scheme. 

\subsection{Baryonic content}\label{sec:baryon}

\begin{figure*}
	\includegraphics[width=0.8\textwidth]{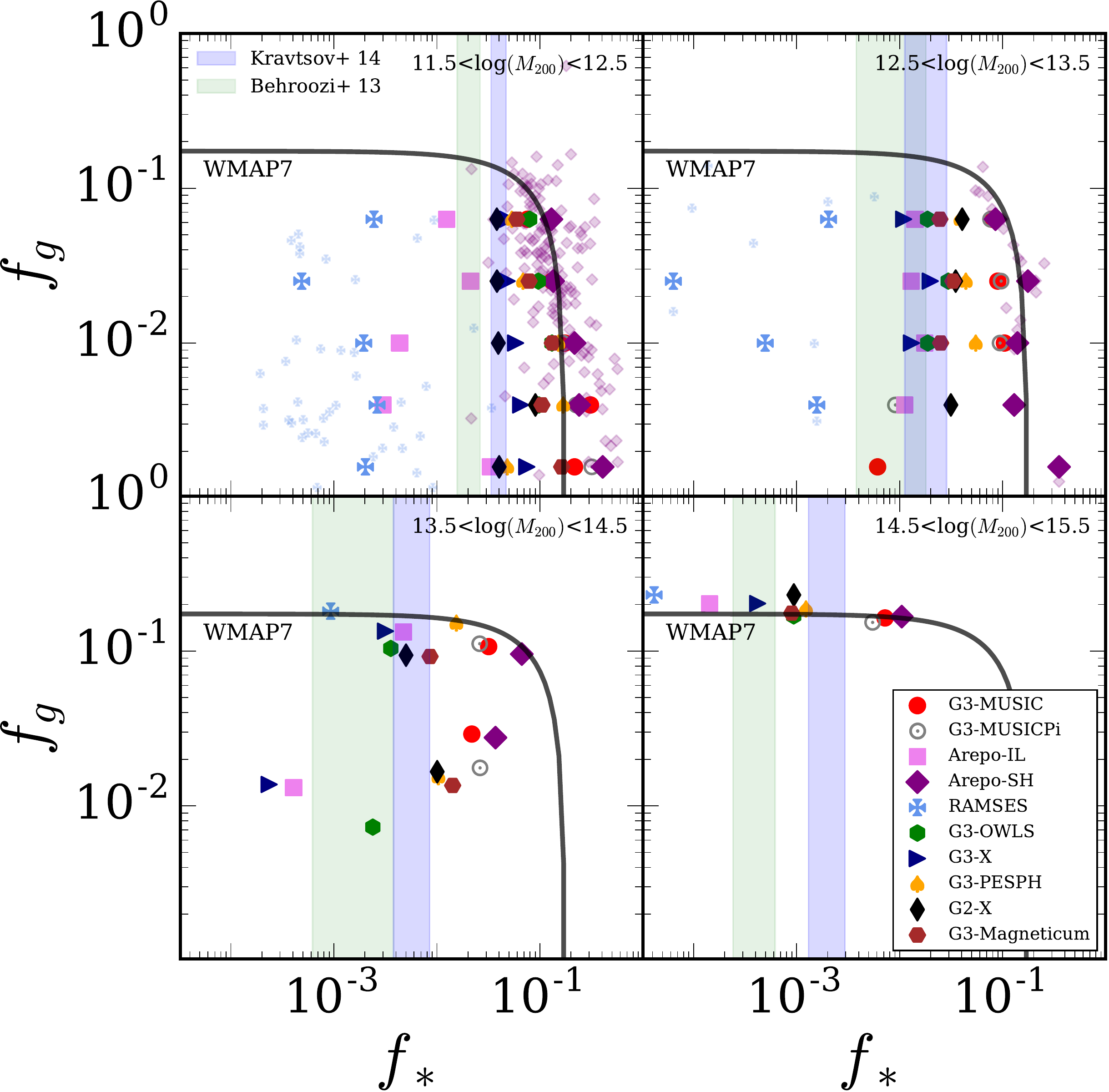}
    \caption{Gas fraction versus stellar fraction in four different
      halo mass ($\text{M}_{200}$) bins as indicated in each panel. The
      highest mass bin, bottom right, shows the baryonic content contained within the main central halo. The bottom left panel indicates the bayonic content for the two most massive sub\halos\ in the main halo. In the top
      panels the large markers represent the average stellar fraction in five
      gas fraction bins, whilst the small markers show the true distribution for the two most extreme cases, \ramses\ and \arepoSH. The cosmic baryon fraction
      ($\Omega_b/\Omega_m$) is shown in each panel as a solid grey
      curve. The green and blue shaded regions represent observational constraints from
      \citet{behroozi13} and \citet{kravtsov2014} respectively. The limits of each observational patch are simply the allowed upper and lower limit in stellar fraction in each mass bin found from each trend. For each code the \halos\ tend to lie in vertical
      bands of stellar fraction and this rank ordering is roughly preserved with mass.}
    \label{fig:gasfracvsstarfrac}
\end{figure*}

\begin{figure}
    \centering
    \includegraphics[width=\columnwidth]{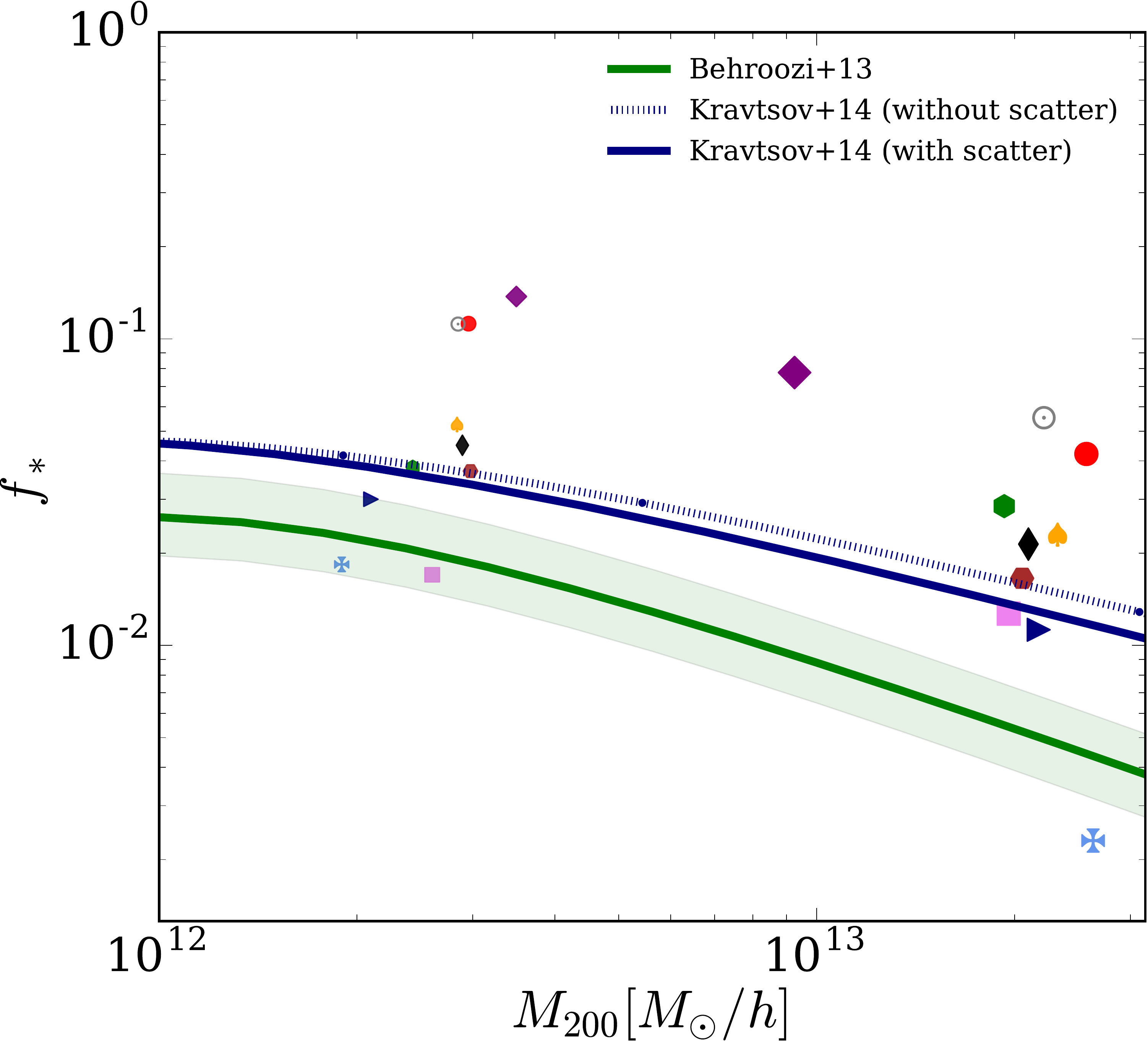}
    \caption{A one-to-one comparison of stellar fraction vs halo mass
      for the two isolated \halos\ marked in \Fig{fig:clusterregion} that are common to all the
      simulations. The smaller and larger markers correspond to a
      $\sim 10^{12}h^{-1}\text{M}_{\odot}$ and $\sim
      10^{13}h^{-1}\text{M}_{\odot}$ mass halo respectively, which are
      both shown as black squares in the bottom right panel of
      \Fig{fig:clusterregion}. Observational constraints from \citet{behroozi13} and \citet{kravtsov2014} are also shown as indicated by the legend. Even
      for both of these relatively isolated \halos\ the code-to-code scatter is still above an order of magnitude.}
    \label{fig:onetoone}
\end{figure}

In order to further investigate what impact the different subgrid prescriptions have on the cluster centre and infall regions, we next study the baryonic material contained within the \halos. In \Fig{fig:gasfracvsstarfrac} we show the gas fraction versus stellar fraction of all \halos\ and sub\halos\ contained within the entire $5h^{-1}\text{Mpc}$ region. We have split these \halos\ into four different mass ($\text{M}_{200}$) bins, shown as different panels in the figure. Observational constraints have also been plotted in each mass bin. The cosmic baryon fraction from WMAP7 data in \citet{komatsu2011} is plotted as a dark grey curve. 

Observed stellar fractions in each mass bin from halo abundance matching relations in \citet{behroozi13} and \citet{kravtsov2014} are shown as green and blue patches respectively. The limits of the patches show the minimum and maximum points in stellar fraction from these trends in each mass range, and are therefore largely exaggerated. Each halo abundance matching trend is derived from a different set of stellar mass functions, which causes some discrepancy between the two, especially in the largest mass bin. The reason why this discrepancy is so large in the largest halo bin is because the stellar mass function \citep{bernardi2013} used in \citet{kravtsov2014} employs an improved photometric method that accounts for the extended stellar envelope surrounding the central BCG. This would lead one to assume that the \citet{kravtsov2014} relation is better suited for modelling galaxy clusters.

This raises the important point that when models calibrate their stellar fraction in the main halo to observational data, they should not include all stellar material contained within the halo as the observations do not account for this. For instance, in this paper we calculate stellar fractions within a sphere of radius $30 h^{-1}\text{kpc}$ centred on the centre of mass of each halo. For all \halos\ except the main cluster halo the differences in simulated stellar fractions between the $30 h^{-1}\text{kpc}$ or whole halo apertures is low ($\lesssim 5\%$). However, for the main halo we find that $\sim80\%$ of stars are located outside the $30 h^{-1}\text{kpc}$ aperture and are part of the intracluster light. Some fraction of stars contained within the intracluster light is partly a numerical artifact associated with simulations at this resolution, and how to deal with them when comparing to observations is still a matter of debate which will be explored in more detail in (Cui et al. (in prep.)). For this study we note that using different sensible apertures doesn't affect the stellar fractions dramatically, for instance changing our $30 h^{-1}\text{kpc}$ aperture to $50 h^{-1}\text{kpc}$ equates to a change in stellar fraction of only $\lesssim 10\%$. In this study we are not worried about this discrapancy as we are not comparing the codes to strict observational limits, as even the two trends included in this paper are in tension in certain mass bins.

The bottom right panel is the equivalent to Fig. 1 in paper II, showing the baryon fraction for the central cluster halo but considering baryonic material within $\text{M}_{200}$ instead of $\text{M}_{500}$. It is clear from this panel that several codes do not reproduce observed stellar fractions. \music, \musicP\ and \arepoSH\ create too many stars by nearly one order of magnitude in the centre compared to observations, which again is not surprising as these codes do no contain AGN feedback. As mentioned before, it is difficult to suggest robust allowed regions of stellar fractions, as even the observations are discrepant by 0.5 dex in this mass bin, but the codes should ideally be aiming to be broadly consistent with at least one set of observations. \ramses\ drastically underproduces stars compared to the observations by $\sim 1$ orders of magnitude. For this single halo, \magneticum, \pesph, \owls, \gxx\ and \gx\ produce stellar fractions that lie between the observations. The bottom left panel indicates where the next two largest \halos\ lie on this plane. As already discussed, these are both within $R_{200}$ and so they are sub\halos\ of the main halo. The code-to-code scatter extends above 2 dex here in stellar fraction.

Interestingly, the ordering of the codes in stellar fraction seen in the bottom panels remains at lower masses where the objects are largely \halos\ in the infall region. Here the large symbols show the average stellar fraction in each gas fraction bin, whilst the small transparent symbols show the scatter for the two most extreme codes, \ramses\ and \arepoSH. Averaged over many \halos\ \owls, \gx, \magneticum\ and \arepoIL\ produce stellar fractions that are more consistent with observations at lower halo masses. Again \ramses\ does not create enough stars by $\sim 1-2$ orders of magnitude. At these masses, we expect the inability of \ramses\ to resolve low mass \halos\ to seriously inhibit its ability to reproduce observed stellar fractions. Again, the stellar fractions for the two \music\ variants and \arepoSH\ are too high, deliberately in the case of \arepoSH\ as this simulation was included to demonstrate the difference turning off AGN feedback made.

The conservation of code ordering in stellar fraction between all four panels again suggests that the primary driver of the scatter is the various subgrid physics implementations, rather than any environmental differences in gas between the codes. We investigated this further by studying the stellar fraction of two specific matched \halos\ in the infall region, marked as black squares in the bottom left panel of \Fig{fig:clusterregion}. We chose these two \halos\ because they were common to all simulations and because they are relatively isolated, so we expect the local gas environments to be more consistent between the models. In this case, isolation means that the \halos\  are well separated from any comparable or larger halo. For instance the two \halos\ have masses $\sim 10^{12}h^{-1}\text{M}_{\odot}$ and $\sim 10^{13}h^{-1}\text{M}_{\odot}$, and the distance from these objects to any other objects with the same mass or above is $\sim 2.4h^{-1}\text{Mpc}$ and $\sim 3.2h^{-1}\text{Mpc}$ respectively. The \halos\ were matched between models by using their halo position and mass.

\begin{figure}
    \centering
    \includegraphics[width=\columnwidth]{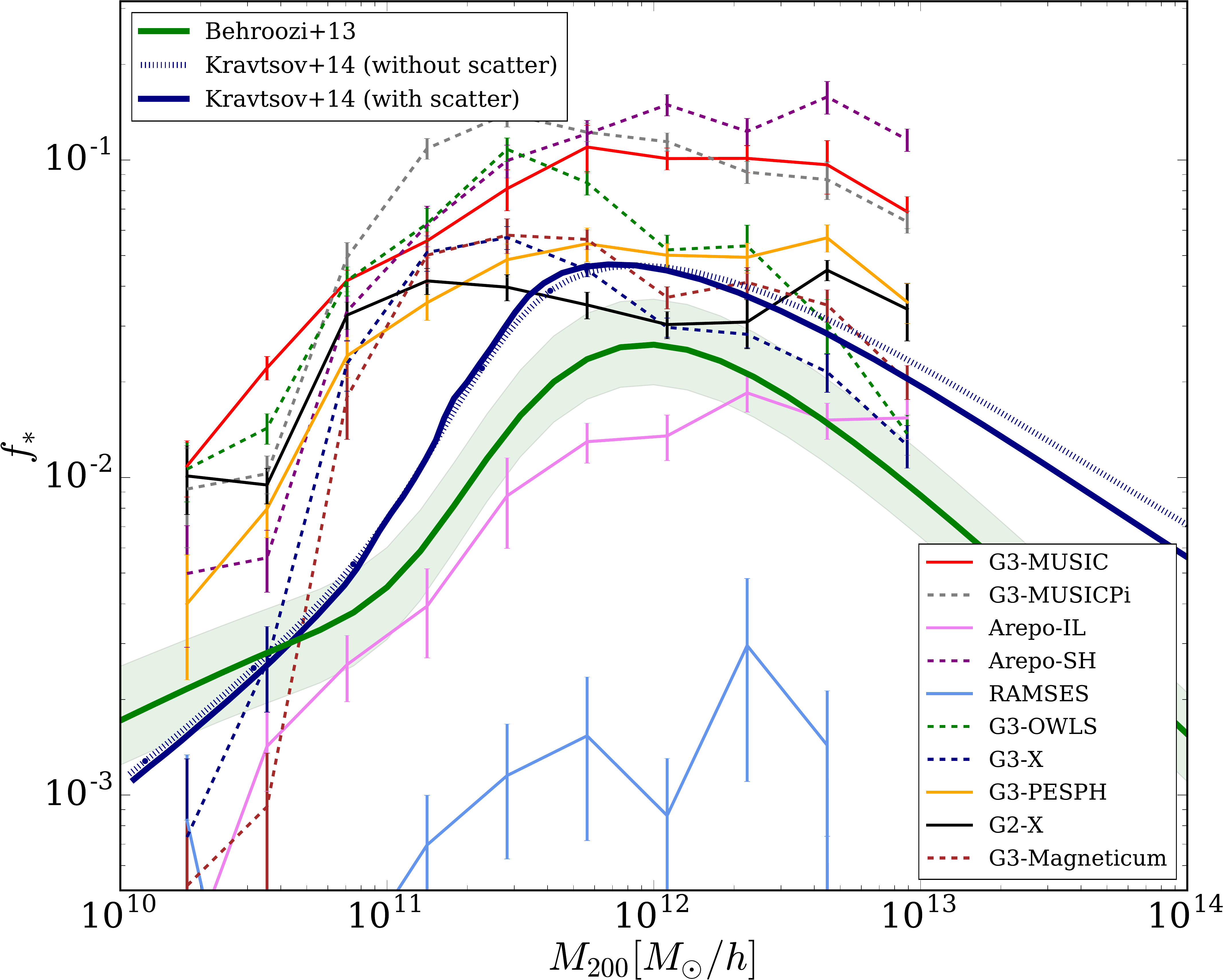}
    \caption{Stellar fraction versus halo mass for each code in the
      cluster infall region. The average stellar
      fractions in each $M_{200}$ bin are presented along with $1\sigma$ errors from the mean obtained from bootstrap sampling. Again observational constraints from \citet{behroozi13} and \citet{kravtsov2014} are also shown as indicated by the legend. At low masses nearly all the codes overproduce
      stars. Five codes produce infall \halos\ that contain stellar fractions that are more consistent with observations above $\sim 10^{12}h^{-1}\text{M}_{\odot}$.  \ramses\ underproduces stars at all halo masses.  }
    \label{fig:stellarfrac}
\end{figure}

\begin{figure}
    \centering
    \includegraphics[width=\columnwidth]{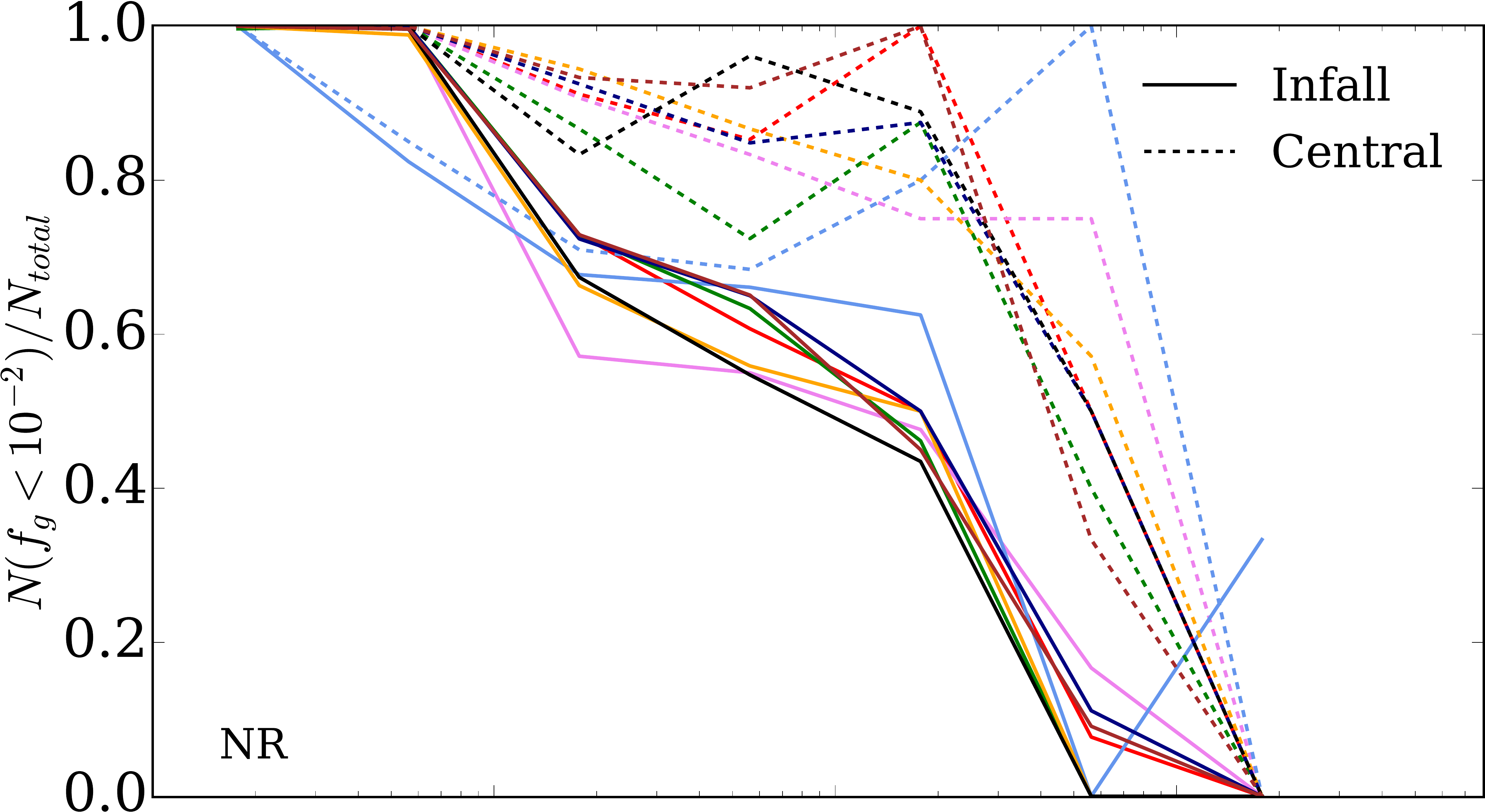}
    \hspace{-2cm}
    \includegraphics[width=\columnwidth]{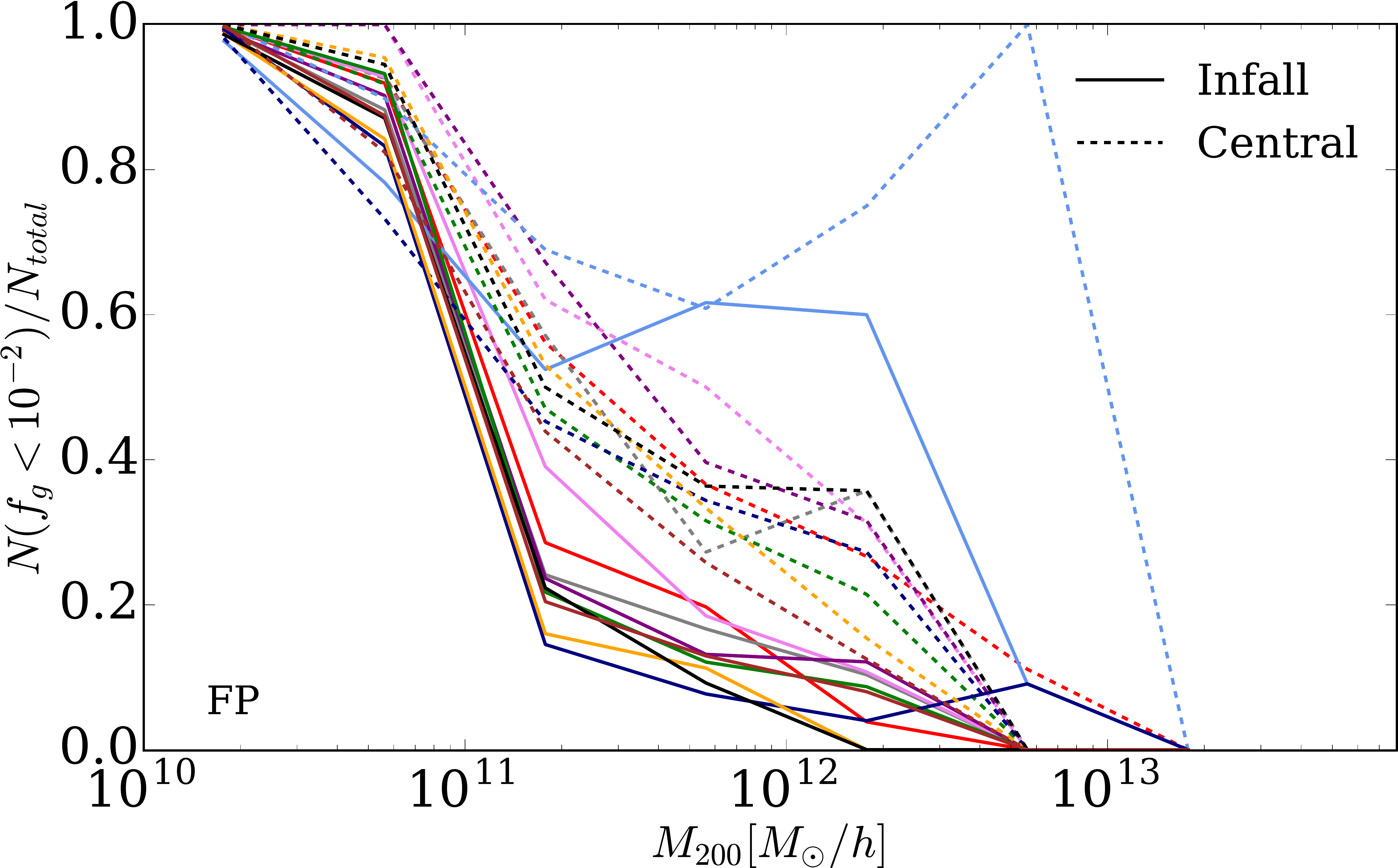}
    \caption{The fraction of gas-poor ($f_g<10^{-2}$) \halos\ as a function halo mass. The non-radiative and full-physics simulations are shown in the top and bottom panels respectively. Solid lines represent \halos\ in the infall region, whilst dashed lines show \halos\ in the central region. See legend in \Fig{fig:cummass} for which coloured line corresponds to which code. In both the central and infall regions, codes produce gas-poor fractions that are typically $\sim50\%$ higher in the NR run compared to the FP run. Codes also tend to produce $\sim20-30\%$ higher gas-poor fractions in the central region compared to the infall region in both the NR and FP runs.}
    \label{fig:gasfrac}
\end{figure}

\Fig{fig:onetoone} shows the stellar fraction vs $M_{200}$ for the isolated \halos\ that are produced by each code, along with the observational constraints from \citet{behroozi13} and \citet{kravtsov2014}. The green shaded regions associated with the \citet{behroozi13} trend are the 1$\sigma$ errors obtained from their MCMC analysis. Two trends are displayed from \citet{kravtsov2014}, one with and one without scatter, where the latter includes artificial scatter when the \halos\ are populated with galaxies in the abundance matching technique. For both \halos\ the code ordering in stellar fraction is again preserved (with small discrepancies) and the code-to-code scatter is still significant. For instance, for the $\sim 10^{12}h^{-1}\text{M}_{\odot}$ halo the difference between the most outlying codes is $0.85$ dex, whilst for the $\sim 10^{13}h^{-1}\text{M}_{\odot}$ halo it is $0.7$ dex, when not including \ramses. Because this amount of scatter is still present in the isolated \halos, we conclude that the differences in the internal  subgrid schemes are driving a large proportion of the code-to-code scatter rather than the different gas environments between the codes.

We have further investigated the stellar fraction versus $\text{M}_{200}$ relation for all \halos\ in the infall region in \Fig{fig:stellarfrac}. Average stellar fractions in $\text{M}_{200}$ mass bins are presented for each code along with $1\sigma$ error bars from the mean obtained from bootstrap sampling. Apart from \arepoIL\ and \ramses, all codes overproduce stars by $\sim 0.1 - 0.6$ dex below $\text{M}_{200} \sim 10^{11.25}h^{-1}\text{M}_{\odot}$. \owls, \magneticum, \gxx, \gx\ and \arepoIL\ produce stellar fractions that are more consistent with either one set of the observations above $\text{M}_{200} \sim 10^{12}h^{-1}\text{M}_{\odot}$. \ramses\ does not produce enough stars by an order of magnitude compared to the \citet{behroozi13} trend across all mass ranges. This figure is troubling, as in cluster simulations it is imperative that all codes are able to match observed stellar fraction vs $\text{M}_{200}$ relations, especially in the infall region as these \halos\ will eventually go on to build the central halo. This issue is becoming increasingly important as galaxy cluster simulations are now being used more widely for cluster cosmology validation (e.g. \citet{mccarthy16}) and environmental galaxy quenching studies (e.g. \citet{bahe2015}).  

We end our analysis with \Fig{fig:gasfrac} where we investigate the fraction of gas-poor ($f_g<10^{-2}$) \halos\ at $z=0$ in each code as a function of halo mass in the infall (solid lines) and central (dashed lines) regions. This allows us to investigate the differences in gas environments between each code in both regions, and to find out which mechanisms may be driving gas out of the \halos. We have done this for both the non-radiative and full-physics runs, shown in the top and bottom panels respectively.

The NR run contains a higher fraction of gas-poor \halos\ compared to the FP run. Above $\sim 10^{11}h^{-1}\text{M}_{\odot}$ where \halos\ are more resolved (\halos\ below this mass contain $<100$ particles), the codes in the NR run produce gas-poor fractions that are typically $\sim50\%$ larger than their FP counterparts. However, with the inclusion of star formation and feedback processes in the FP run, we would naively expect there to be a higher gas-poor fraction here. Presumably this means that either the extra gravitional pull from the stars is enough to retain the gas or that the employed feedback schemes are not powerful enough to drive outflows, which could be linked to the overcooling problems seen in \Fig{fig:gasfracvsstarfrac}. However, the reason for the discrepancy between the NR and FP runs could be that the gas in the NR run cannot cool, unlike in the FP run. Therefore, the gas may remain extended in the NR run and more easily stripped.

There are also differences in the gas-poor fractions between the central and infall regions. In both the NR and FP run, codes in the central region typically produce gas-poor fractions that are $\sim20-30\%$ larger than the infall region. We expect the differences in the gas-poor fractions between the central and infall regions to be predominantly due to the gas in the \halos\ being more efficiently stripped in the centre by the increased ram pressure.

\section{Discussion and Conclusions}\label{sec:conclusions}

Hydrodynamical simulations of galaxy clusters are now vital tools for interpreting and understanding observational data. However, it is vital that the validity of the models used to produce such simulations are checked by carrying out model comparison studies. This paper is a continuation of one such study, the \textit{nIFTy cluster comparison  project} whose aim is to take eight state-of-the-art hydrodynamical codes each equipped with their own calibrated subgrid physics and to examine a $\text{M}_{200} = 1.1\times10^{15}h^{-1}\text{M}_{\odot}$ galaxy cluster each model produces from the same initial conditions.

In this paper we have studied the properties of \halos, sub\halos\ and galaxies residing in the infall region ($\text{R}_{200}-5h^{-1}\text{Mpc} (\sim 3R_{200})$) surrounding this cluster. This is an extension of the work done in \citet{elahi2015} (E16) who carried out a similar study inside $\text{R}_{200}$ of the same synthetic cluster, where they found striking code-to-code differences in galaxy abundances and mass.

We have studied how well each model reproduces observed stellar fraction vs halo mass relations, further investigated the sources of code-to-code scatter seen in \citet{elahi2015} and examined the extent to which ongoing preprocessing is occuring in the infall region at $z=0$. Our main conclusions are presented below along with some discussion.

\begin{itemize}
 \item We have presented the $\text{M}_{200} = 1.1\times10^{15}h^{-1}\text{M}_{\odot}$ nIFTy galaxy cluster showing the dark matter, gas and stellar content along with the halo distribtution in the infall region. It is clear that the galaxy cluster is surrounded by obvious filamentary structure that hosts 2-3 group sized ($>10^{13}h^{-1}\text{M}_{\odot}$) halos.
\item After comparing the number of \halos\ and sub\halos\ between codes in the infall region, we have found that although there is more scatter in the Full-Physics (FP) run compared to the Dark-Matter (DM) only and Non-Radiative (NR) runs, the code-to-code scatter is still $<15\%$. The exception is the AMR code \ramses, which produces a factor of two fewer \halos\ and sub\halos\ than the median. Along with an over-powered AGN feedback scheme, this is partly a resolution issue that is inherent to AMR codes as \ramses\ is more aligned with other codes for \halos\ containing 200 dark matter particles or more.
\item The code-to-code scatter in galaxy abundance in the central region seen in E16 extended up to a factor $\sim20$ between the two most extreme cases. We have shown that the same degree of scatter is still present in the infall region as well, which suggests that the code-to-code scatter seen in E16 is predominantly due to the different subgrid implementations employed by each code, rather than any differences in gas environments between the codes, which would be exacerbated in the overdense central region compared to the infall region. Codes without AGN feedback such as \music, \musicP\ and \arepoSH\ produce the most galaxies, whilst \ramses\ and \arepoIL\ produce the least.
\item In all codes we have shown that there are $\sim10$ times more \halos\ than sub\halos\ in the infall region, which is as expected from dark-matter-only simulations (e.g. \citet{klypin2011}). The small subhalo to halo ratio suggests that there may not be much ongoing preprocessing at $z=0$, which would be in tension with recent observational studies that have suggested preprocessing is a dominate mechanism in the infall region at $z\sim0$ \citep{cybulski2014,just2015}. However, we caution that this may not be a fair comparison, and we intend to carry out a full temporal study in order to investigate preprocessing in the infall region as this cluster forms. 
\end{itemize}
\begin{itemize}
\item We also compared estimates of halo mass and maximum circular velocity, which has been suggested as a better statistic from which to derive mass. We notice a significant increase in code-to-code scatter in the measurement of the maximum circular velocity for large \halos\ in the FP models compared to the $\text{M}_{200}$ estimate. This is because the maximum circular velocity occurs close to the halo centre and this region is significantly disturbed by the feedback schemes employed in the FP run. We caution that the use of maximum circular velocity may not lead to the significant improvement suggested for FP models.
\item We have shown that five codes do not reproduce observed stellar fractions \citep{behroozi13,kravtsov2014} for the main cluster halo, typically the ones not containing AGN feedback that overproduce stars, as well as \arepoIL\ and \ramses, which underproduce stars compared to observations. For this halo, the scatter in stellar fraction between the two most extreme codes is around two orders of magnitude. Averaged over many \halos\ the story is the same at lower halo masses, where the same degree of code-to-code scatter is still present and the rank ordering of codes in stellar fraction is roughly preserved. \gx\ and \owls\ are the most consistent with observations in all mass bins in \Fig{fig:gasfracvsstarfrac}. However, we do caution that the two observational trends used in this study are in tension with each other, due to the different set stellar mass functions each uses to produce their relations. Though we expect the \citet{kravtsov2014} relation to be more suitable at the high mass end due to its use of a stellar mass function \citep{bernardi2013}, which uses an improved photometric method to capture the outer envelope of the cluster BCG. 
\item After analysing the stellar fractions of two isolated \halos\ (with mass $\sim 10^{12}h^{-1}\text{M}_{\odot}$ and $\sim 10^{13}h^{-1}\text{M}_{\odot}$) common to all models in the infall region, we find that the code-to-code scatter is still above $>1$ dex for both objects. As these \halos\ are far enough away from any neighbouring \halos\ of comparable mass ($>2h^{-1}\text{Mpc}$), we expect this scatter to be predominantly due to the differences in the internal subgrid implementations rather than any external gas environment differences between the models.
\item By comparing the stellar fraction vs $\text{M}_{200}$ of all \halos\ only in the infall region to observed trends from \citet{behroozi13} and \citet{kravtsov2014}, we find that \owls, \magneticum, \gxx, \pesph, \gx\ and \arepoIL\ are reasonably consistent with either set of observations above $\sim 10^{11.25}h^{-1}\text{M}_{\odot}$ (differences between models and observations here is typically $\lesssim 0.2$ dex). Below this mass all of the \gadget\ variant models produce too many stars compared to the observed stellar fractions by nearly an order of magnitude, which is presumably a resolution issue as these \halos\ will only contain $\lesssim 100$ particles. This issue is hard to solve as it is often unfeasible to produce massive galaxy cluster simulations with better resolution than in our study. \music, \musicP, \arepoSH\ and \ramses\ are discrepant with observations at all halo masses by $\gtrsim 0.5$ dex, because they either do not contain AGN feedback (deliberately in the case of \arepoSH) or in the \ramses\ case the AGN is far too powerful.
\item The inability of \ramses\ to reproduce observations by consistently underproducing stellar material within \halos\ and sub\halos\ of every mass is in stark tension with the Rhapsody-G simulations studied in \citet{hahn2015}. They studied ten galaxy clusters simulated with \ramses\ of similar mass and resolution to the nIFTy cluster, and found good agreement between the stellar content contained within the \halos\ and sub\halos\ surrounding the clusters to halo abundance matching trends. We suspect the differences between these two results to arise from the fact that \ramses\ includes variant subgrid prescriptions between the two runs that have been calibrated differently. Many subgrid models can be calibrated to repoduce different targeted observables, but this doesn't necessarily mean one is more accurate or reliable than the other. These subgrid models are simply recipes with knobs that can be turned in order to reproduce specific things, and one cannot disregard one code because it does not match one key observable. 
\end{itemize}

In the future we expect these codes and many more to continously improve by incorporating more realistic subgrid models that are extensively calibrated to current and new observables (e.g. \citet{mccarthy16}) at $z=0$ and above, which in turn will lead to more accurate cluster simulations from which valuable science can be done. We next intend to carry out a full temporal study within a larger $25h^{-1}\text{Mpc}$ zoom region surrounding this cluster in order to investigate the assembly history of the cluster and the effectiveness of preprocessing at higher redshift.

\section*{Acknowledgments}
The authors would like the acknowledge the Centre for High Performance Computing in Rosebank, Cape Town for financial support and for hosting the "Comparison Cape Town" workshop in July 2016. The authors would further like to acknowledge the support of the International Centre for Radio Astronomy Research (ICRAR) node at the University of Western Australia (UWA) in the hosting the precursor workshop "Perth Simulated Cluster Comparison" workshop in March 2015; the financial support of the UWA Research Collaboration Award 2014 and 2015 schemes; the financial support of the ARC Centre of Excellence for All Sky Astrophysics (CAASTRO) CE110001020; and ARC Discovery Projects DP130100117 and DP140100198. We would also like to thank the Instituto de Fisica Teorica (IFT-UAM/CSIC in Madrid) for its support, via the Centro de Excelencia Severo Ochoa Program under Grant No. SEV-2012-0249, during the three week workshop ``nIFTy Cosmology" in 2014, where the foundation for the whole comparison project was established.

JTA acknowledges support from a postgraduate award from STFC. PJE is supported by the SSimPL programme and the Sydney Institute for Astronomy (SIfA), and {\it Australian Research Council} (ARC) grants DP130100117 and DP140100198. AK is supported by the {\it Ministerio de Econom\'ia y Competitividad} (MINECO) in Spain through grant AYA2012-31101 as well as the Consolider-Ingenio 2010 Programme of the {\it Spanish Ministerio de Ciencia e Innovaci\'on} (MICINN) under grant MultiDark CSD2009-00064. He also acknowledges support from the {\it Australian Research Council} (ARC) grant DP140100198. He further thanks Noonday Underground for surface noise. STK acknowledges support from STFC through grant ST/L000768/1. CP acknowledges support of the Australian Research Council (ARC) through Future Fellowship FT130100041 and Discovery Project DP140100198. 
WC and CP acknowledge support of ARC DP130100117. GY and FS acknowledge support from MINECO (Spain) through the grant AYA 2012-31101. GY thanks also the  Red Espa\~{n}ola de Supercomputacion for granting the computing time in the Marenostrum Supercomputer at BSC, where all the MUSIC simulations have been performed. AMB is supported by the DFG Research Unit 1254 ``Magnetisation of interstellar and intergalactic media'' and by the DFG Cluster of Excellence ``Universe''. SB \& GM acknowledge support from the PRIN-MIUR 2012 Grant "The Evolution of Cosmic Baryons" funded by the Italian Minister of University and Research, by the PRIN-INAF 2012 Grant "Multi-scale Simulations of Cosmic Structures", by the INFN INDARK Grant and by the "Consorzio per la Fisica di Trieste". IGM acknowledges support from a STFC Advanced Fellowship. EP acknowledges support by the ERC grant ``The Emergence of Structure during the epoch of Reionization''.

The authors contributed to this paper in the following ways: JTA, FRP, MEG, PJE and AK formed the core team. JTA analysed the data, made the plots and wrote the paper. FRP and MEG assisted in writing the paper. PJE assisted with the analysis. AK, GY and FRP organised the nIFTy workshop. GY supplied the initial conditions. All other authors performed simulations using their codes or read and commented on the paper.

The simulations used for this paper have been run on a variety of supercomputers and are publicly available at the MUSIC website, \href{http://www.music.ft.uam.es}{\url{http://www.music.ft.uam.es}}. MUSIC simulations were carried out on Marenostrum. AREPO simulations were performed with resources awarded through STFCs DiRAC initiative. The authors thank Volker Springel for helpful discussions and for making AREPO and the original GADGET version available for this project. The authors also thank Andrey Kravtsov for useful discussions. G3-PESPH Simulations were carried out using resources at the Center for High Performance Computing in Cape Town, South Africa. 


\bibliographystyle{mnras}
\bibliography{comparison_outskirts_v2} 

\bsp	
\label{lastpage}
\end{document}